\documentclass[reprint,superscriptaddress,twocolumn, aps, prb]{revtex4-1}

\usepackage{graphicx,psfrag}
\usepackage{epstopdf}
\usepackage{amsmath} 
\usepackage{bm}
\usepackage{amssymb}
\usepackage{subfigure}
\usepackage{color}
\usepackage[pagewise]{lineno}
\usepackage[normalem]{ulem}

\usepackage{titlesec}
\begin{document}

\title{Occurrence of anomalous diffusion and non-local response in highly-scattering acoustic periodic media}

\author{Salvatore Buonocore}
\author{Mihir Sen}
\affiliation{Department of Aerospace and Mechanical Engineering, University of Notre Dame, Notre Dame, IN 46556}
\author{Fabio Semperlotti}
\email{To whom correspondence should be addressed: fsemperl@purdue.edu}
\affiliation{Ray W. Herrick Laboratories, School of Mechanical Engineering, Purdue University, West Lafayette, IN 47907}

\begin{abstract}

We investigate the occurrence of anomalous diffusive transport associated with acoustic wave fields propagating through highly-scattering periodic media. Previous studies had correlated the occurrence of anomalous diffusion to either the random properties of the scattering medium or to the presence of localized disorder. In this study, we show that anomalous diffusive transport can occur also in perfectly periodic media and in the absence of disorder.  
The analysis of the fundamental physical mechanism leading to this unexpected behavior is performed via a combination of deterministic, stochastic, and fractional-order models in order to capture the different elements contributing to this phenomenon. Results indicate that this anomalous transport can indeed occur in perfectly periodic media when the dispersion behavior is characterized by anisotropic (partial) bandgaps. In selected frequency ranges, the propagation of acoustic waves not only becomes diffusive but its intensity distribution acquires a distinctive L{\'e}vy $\alpha$-stable profile having pronounced heavy-tails. In these ranges, the acoustic transport in the medium occurs according to a hybrid transport mechanism which is simultaneously propagating and anomalously diffusive. We show that such behavior is well captured by a fractional diffusive transport model whose order can be obtained by the analysis of the heavy tails.

\end{abstract}

\maketitle

\section{Introduction}  \label{Introduction}

In recent years, several theoretical and experimental studies have shown that field transport processes in non-homogeneous and complex media can occur according to either hybrid or anomalous mechanisms. Some examples of these physical mechanisms include anomalous diffusive transport (such as non-Fourier \cite{povstenko2013fractional,borino2011non,ezzat2010thermoelectric}, or non-Fickian diffusion \cite{ benson2000application,benson2001fractional,cushman2000fractional,fomin2005effect} with heavy-tailed distribution) or hybrid wave transport (characterized by simultaneous propagation and diffusion 
\cite{mainardi1996fractional,mainardi1996fundamental,mainardi1994special,mainardi2010fractional,chen2003modified,chen2004fractional}). Simultaneous hybrid and anomalous transport has also been observed, particularly in wave propagation problems involving random scattering media. Electromagnetic waves traveling through a scattering material\cite{yamilov2014position} such as fog \cite{belin2008display} or murky water \cite{zevallos2005time} are relevant examples of practical problems where such transport process can arise. 

A distinctive feature of anomalous transport is the occurrence of heavy-tailed distributions of the representative field quantities \cite{benson2001fractional}. In this case, the diffusion process does not follow a classical Gaussian distribution but instead is characterized by a high-probability of occurrence of the events associated with large variance (i.e. those described by the "heavy" tails).

This behavior is typically not accounted for in traditional field transport theories based on integer order differential or integral models. Purely numerical methods, such as Monte Carlo or finite element simulations\cite{huang1991optical,ishimaru2012imaging,mosk2012controlling,sebbah2012waves,gibson2005recent}, can capture this response but are very computationally intensive and do not provide any additional insight in the physical mechanisms generating the macroscopic dynamic behavior. The ability to accurately predict the anomalous response and to retrieve information hidden in diffused fields remains a challenging and extremely important topic in many applications. Acoustical and optical imaging, non-intrusive monitoring of engineering and biomedical materials are just a few examples of practical problems in which the ability to carefully predict the field distribution is of paramount importance to achieve accurate and physically meaningful solutions. Nevertheless, in most classical approaches, information contained in the heavy tails is typically discarded because it cannot be properly captured and interpreted by integer-order transport models.
 
Hybrid and anomalous diffusive transport mechanisms are pervasive also in acoustics. This type of transport can arise when acoustic fields propagate in a highly scattering medium such as a urban environment \cite{albert2010effect,remillieux2012experimental}, a forest \cite{aylor1972noise,tarrero2008sound}, a stratified fluid (e.g. the ocean) \cite{baggeroer1993overview,dowling2015acoustic,casasanta2012fractional}, or a porous medium \cite{benson2001fractional,schumer2001eulerian,fellah2003measuring,fellah2000transient}.   
		
From a general perspective, classical diffusion of wave fields occurs within a range where the wavelength is comparable to the size of the scatterers, the so-called Mie scattering regime. Any deviation from classical diffusion, being either sub-diffusion \cite{metzler2000random,goychuk2012fractional}  (typically linked to Anderson
localization) or super-diffusion (typically linked to L{\'e}vy-flights) \cite{barthelemy2008levy,bertolotti2010multiple}, still arises within the same regime. The two dominant factors are either the relation between the transport mean free path and the wavelength, or the statistical distributions of the scattering paths in presence of disorder. 
When a wave field interacts with scattering elements, it undergoes a variety of physical phenomena including reflection, refraction, diffraction, and absorption that significantly alter its initial characteristics. Depending on the quantity, distribution, and properties of the scatterers the momentum vector of an initially coherent wave can become quickly randomized.
For most processes, the Central Limit Theorem (CLT) guarantees that the distribution of macroscopic observable quantities (e.g. the field intensity) converges to a Gaussian profile in full agreement with the predictions from classical Fourier diffusion. At the same time, the transition to a macroscopic diffusion behavior leads to an inevitable coexistence of diffusive and wave-like processes at the meso- and macro-scales. 

There are numerous physical processes in nature whose \textit{basin of attraction} is given by the normal (Gaussian) distribution. On the other hand, when the distribution of characteristic step-length has infinite variance, the diffusion process no longer follows the standard diffusion theory, but rather acquires an anomalous behavior with a basin of attraction given by the so-called $\alpha$-stable L{\'e}vy distribution. In the latter case, the unbounded value of the variance of the step-length distribution is due to the non-negligible probability of existence of steps whose lengths greatly differ from the mean value; these are usually denoted as L{\'e}vy flights. The distinctive feature of the $\alpha$-stable L{\'e}vy distributions is the occurrence of heavy tails having a power-law decay of the form $p(l) \sim l^{-(\alpha+1)}$. This characteristic suggests that transport phenomena evolving according to L{\'e}vy statistics are dominated by infrequent but very long steps, and therefore their dynamics are profoundly different from those predicted by the random (Brownian) motion. 
Many of the complex hybrid transport mechanisms mentioned above fall in this category, and therefore cannot be successfully described in the framework of classical diffusion theory.
 
In addition, these complex transport mechanisms are typically not amenable to closed-form analytical solutions therefore requiring either fully numerical or statistical approaches to predict the field quantities under various input conditions. Typical modeling approaches rely on random walk statistical models \cite{metzler2000random,bouchaud1990anomalous} or on semi-empirical corrections to the fundamental diffusive transport equation via renormalization theory \cite{asatryan2003diffusion,cobus2016anderson}. These approaches imply a considerable computational cost and do not provide physical insight in the operating mechanisms of the anomalous response.
A few studies have also indicated that, for this type of processes, the macroscopic governing equation describing the evolution of the wave field intensity could be described by a generalization of the classical diffusion equation using fractional derivatives
\cite{bertolotti2010multiple,metzler2000random,bertolotti2007light}.

To-date, the occurrence of anomalous diffusion of wave fields has been connected and observed only in random and disordered media \cite{barthelemy2008levy,burresi2012weak,bouchaud1990anomalous,asatryan2003diffusion, cobus2016anderson}. In this study, we show theoretical and numerical evidence that anomalous behavior can occur even in presence of perfectly periodic media and in absence of disorder. We present this analysis in the context of diffusive transport of acoustic waves although the results could be generalized to other wave fields.
In particular, we investigate the specific case of propagation of acoustic waves in a medium with identical and periodically distributed hard scatterers.
We develop a theoretical framework for multiple scattering in super-diffusive periodic media. 
We first show, by full field numerical simulations, that under certain conditions, acoustic waves propagating through a periodic medium are subject to anomalous diffusion. Then, we propose an approach based on a combination of deterministic and stochastic methodologies to explore the physical origin of this unexpected behavior. Ultimately, we show that fractional order models can predict, more accurately and effectively, the resulting anomalous field quantities. More important, we will show that the analysis of the heavy tails provide a reliable means to extract the equivalent fractional order of the medium. 

\section{Anomalous diffusion in acoustic periodic media: overview of the method}
	
\begin{figure}[h!]
\center{\includegraphics[width= 6 cm]{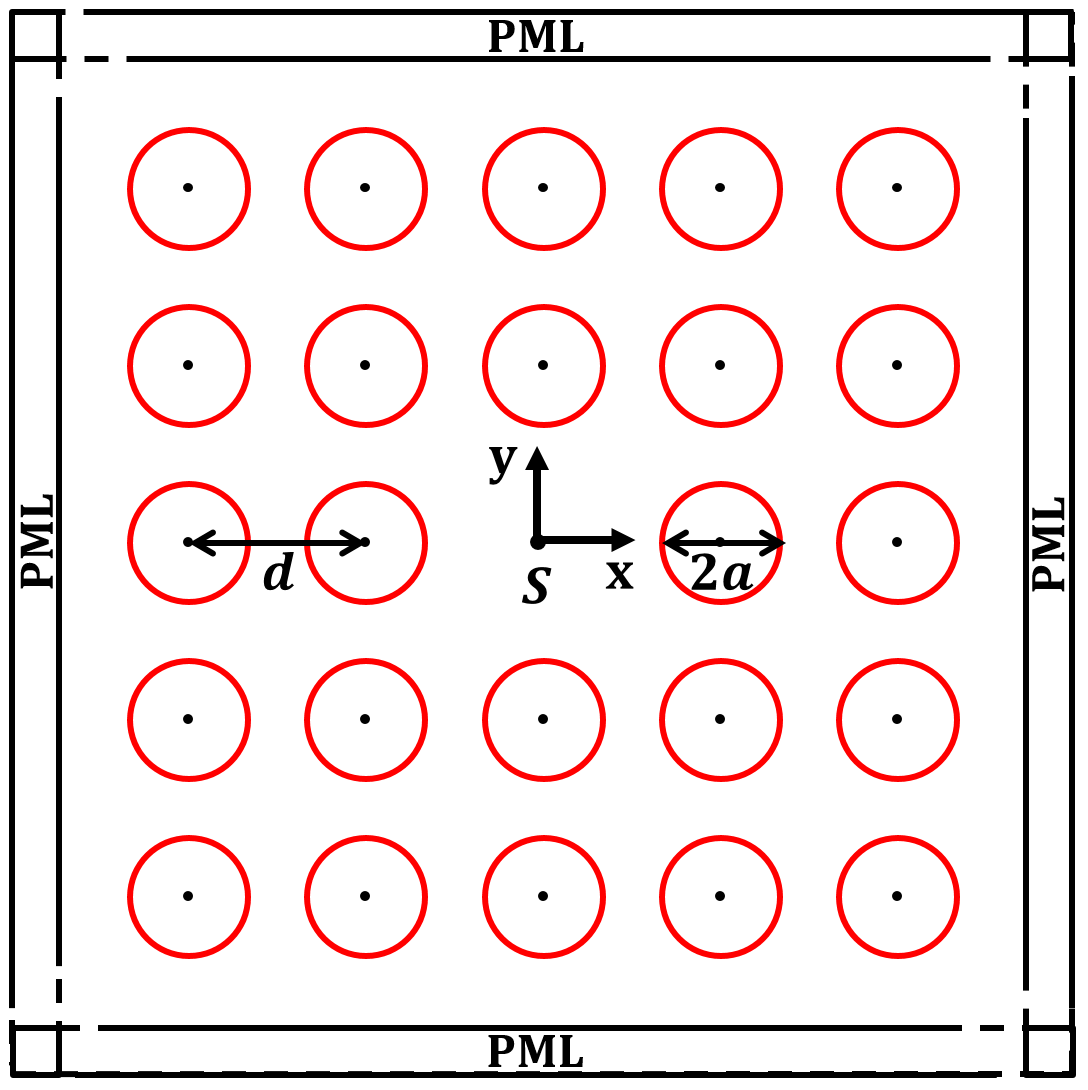}
  \caption{Schematic of a 2D periodic square lattice of hard cylindrical scatterers that will be used to show the occurrence of anomalous hybrid propagation. A monopole-like acoustic source is located in the center as indicated by the label $S$. For the numerical simulation the acoustic medium is enclosed by perfectly matched layers (PML) in order to eliminate reflections from the boundaries.}
\label{Fig_1}}
\end{figure}

We consider the generic problem of an acoustic bulk medium made of periodically-distributed cylindrical hard scatterers in air (Fig.~\ref{Fig_1}). We assume a monopole-like acoustic source, located in the center of the lattice, which emits at a selected harmonic frequency chosen within the scattering regime.

The main objective is to characterize the propagation of acoustic waves in such medium based on different regimes of dispersion. As previously anticipated, in selected regimes the propagation of acoustic waves will exhibit anomalous diffusive transport properties.
The reminder of this study will be dedicated to investigating the causes leading to the occurrence of such phenomenon. In order to identify the fundamental mechanisms at the origin of this behavior, we have designed a multi-folded approach capable of characterizing the different processes contributing to the macrosopic anomalous response. 

The approach consists of the following components.
First, we investigate via full-field numerical simulations the propagation of acoustic waves in either a 1D or a 2D periodic scattering medium. The numerical results will allow making important observations on the different propagation mechanisms occurring in the two systems and on the corresponding diffusive processes.
Then, the radiative transfer theory will be applied to interpret the evolution of the wave intensity distribution and analyze the nature of the diffusive phenomena in the context of a renormalization approach. 

In order to identify the physical mechanism at the origin of the anomalous diffusion, a multiple scattering analysis based on the multipole expansion method will be applied in order to characterize the interaction between different scatterers. In particular, this approach was intended to identify and quantify possible long-range interactions between pairs of scatterers. Based on the results of the multiple scattering analysis, a Monte Carlo model is used to confirm that the anomalous transport is in fact originated by the long-range interactions between different directions of propagation in the lattice. 

Finally, we show that the behavior of the lattice can be effectively described in a homogenized sense, by a fractional continuum diffusion model whose fractional order can be identified by fitting an $\alpha$-stable distribution to the heavy tails of the wave intensity. This approach can be seen as an equivalent \textit{fractional homogenization} of the medium. Of particular interest is the fact that the fractional (homogenized) model allows a closed-form analytical solution the agrees very well with the numerical predictions.

\section{Scattering and diffusive transport}  \label{Overview}

From a general perspective, it is possible to identify four different wave propagation regimes in scattering media which are classified based on the relative ratio of quantities such as the transport mean free path $l_t$, the wavelength of the propagating field $\lambda$, and the characteristic size of the scattering domain $L$. The four regimes are:

\begin{enumerate}
    \item  The \textit{homogenized regime}: it occurs when the wavelength of the incident wave field is much larger than the typical characteristic size $d$ of the scatterer, that is $\lambda \gg d$.

   \item The \textit{diffusive regime}: it occurs when the wavelength $\lambda$ satisfies the relation $\lambda/2\pi \ll l_t \ll L $. 
   In this regime the wave intensity can be approximated by the diffusion equation.

   \item The \textit{anomalously diffusive regime}: it occurs when the interference of waves causes the reduction of the transport mean free path $l_{t}$ and consequently a renormalization of the macroscopic diffusion constant $D$. In this regime, the transport mean free path varies according to the size of the cluster and to the degree of disorder.

   \item The \textit{localization regime}: it occurs in the range $\lambda/2\pi \geq l_{t}$ and corresponds to a diffusion constant $D$ tending to 0.
\end{enumerate}

As mentioned in the classification above, there are regimes in which the intensity of the wave field can be properly described by the \textit{diffusion approximation}, that is it varies in space as prescribed by the field evolution in a diffusion equation. 
In particular, when the incident wave has a wavelength smaller than the length-scale characterizing the material and/or of the geometric variations of the physical medium, the wave field undergoes multiple scattering with a consequent randomization of its phase and direction of propagation. In order to characterize this phenomenon a statistical description based on random walk models is typically employed. These models rely on phenomenological quantities such as the scattering $l_s$ and the transport $l_t$ mean free paths. From a physical perspective, $l_s$ represents the average distance between two successive scattering events, while $l_t$ is the mean distance after which the wave field loses memory of its initial direction and becomes randomized \cite{van1999multiple}. When the filling factor $f$ (which describes the density of scatterers) is low, $l_{s}$ and $l_{t}$ are defined as \cite{ishimaru1978wave}:

 \begin{eqnarray} \label{ls_lt}
l_{s} &=& \dfrac{1}{\rho\sigma_{t}}\\
l_{t} &=& \dfrac{l_{s}}{1-\langle\cos{\theta}\rangle} \nonumber
\label{eq:lt}
\end{eqnarray}

\noindent where $\sigma_{t}$ is the total scattering cross section, $\langle\cos{\theta}\rangle$ is the \textit{anisotropy factor} and $\rho$ is the scatterers concentration. Note that the relations in Eq.~(\ref{ls_lt}) are valid only for low filling factors, approximately in the range $f \leq 0.1$. For increased values of the filling factor, the scattering cross section $\sigma_t$ needs to be rescaled. The rescaling factor in the range $0.1 \leq f \leq 0.6$ is given by $\sigma_t \rightarrow \sigma_t (1-f)$, while higher filling factors require a more elaborated rescaling procedure \cite{ishimaru1978wave}.

The scattering cross section plays a crucial role in the characterization of multiple scattering phenomena and in two dimensions takes the form:

\begin{eqnarray} \label{sigma_t}
\begin{split}
\sigma_t = \int_{2\pi}\sigma_d(\theta) d\theta .
\end{split}
\end{eqnarray}

The integrand $\sigma_d$ is the differential scattering cross section defined as:

\begin{eqnarray} \label{diff_scatt_cross_sect}
\begin{split}
\sigma_d(\theta)= \lim_{R \rightarrow \infty} R\left [ (S_s(\theta))/S_i\right ].
\end{split} 
\end{eqnarray}.

In Eq.~(\ref{diff_scatt_cross_sect}),  the term $S_s$ is the scattered power flux density at a distance $R$ from the scatterer in the direction $\mathbf{\hat{o}}$ caused by an incident power flux density $S_i$. The azimuthal angle $\theta$ is the angle between the incident ($\mathbf{\hat{i}}$) and the scattered wave fields ($\mathbf{\hat{o}}$).

\begin{figure}[h!]
\center{\includegraphics[width= 5 cm]{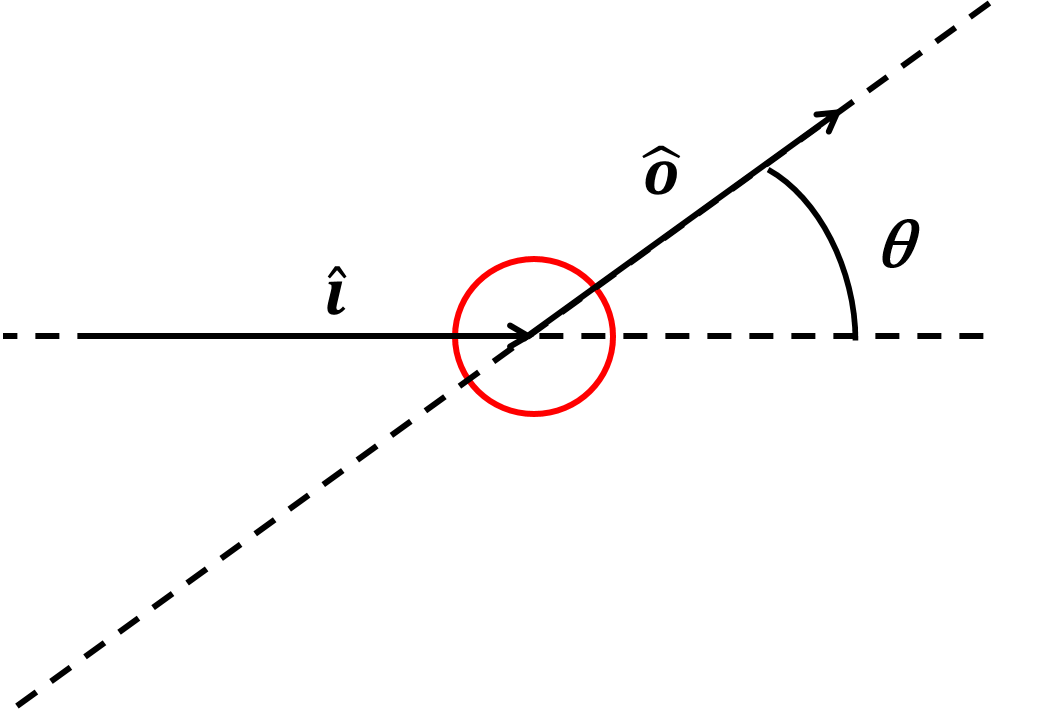}}
  \caption{Schematic of a scattering event of a plane wave due to a cylindrical object. $\mathbf{\hat{i}}$ and  $\mathbf{\hat{o}}$ represent the direction of the incident and refracted wave, respectively.}
\label{Fig_2}
\end{figure}
 
The \textit{scattering phase function} is obtained by normalizing the differential scattering cross section with respect to $\sigma_t$:

\begin{eqnarray} \label{}
p\mathbf{( \hat{o},\hat{i})}=p(\cos\theta) = \frac{\sigma_d(\theta)}{\sigma_{t}}
 \end{eqnarray}
 
\noindent and represents the probability that a wave field impinging on the scatterer from a given direction will be scattered by an angle $\theta$. The mean value of the previous probability distribution defines the \textit{anisotropy factor}:
 
\begin{eqnarray} \label{}
\langle \cos \theta \rangle = \int_{2\pi} p(\cos \theta) \cos(\theta) d \theta.
 \end{eqnarray}
 
This factor varies between 0 and 1, and it accounts for the existence of preferential scattering directions. For $\langle\cos{\theta}\rangle=0$ all the scattering directions have the same probability and the scattering is isotropic. As $\langle\cos{\theta}\rangle$ approaches 1, the forward scattering becomes the most probable event. These quantities will be used in the following analyses in order to identify the different scattering regimes.

\section{One-dimensional medium}
\begin{figure}[h!]
\center{\includegraphics[width= 8 cm]{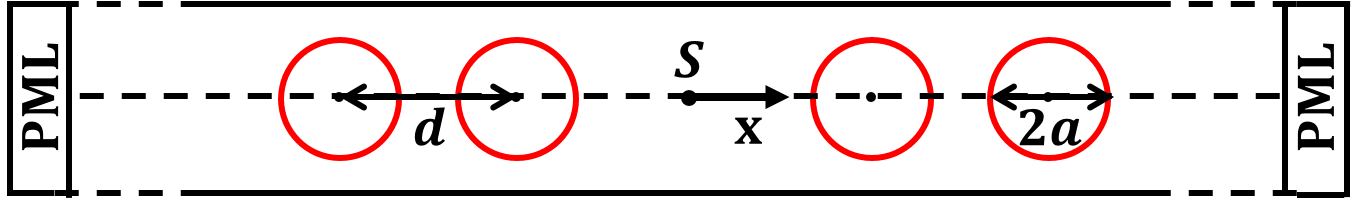}
  \caption{Schematic of the 1D periodic waveguide made of hard cylindrical scatterers in air. A monopole-like acoustic source is located at $S$. Perfectly matched layers (PML) are used to simulate an infinite guide.}
\label{Fig_3}}
\end{figure}

Consider a one-dimensional bulk scattering medium composed of $N$ hard cylindrical scatterers equally distributed in an air background (Fig.~\ref{Fig_3}).

This system can be interpreted by all means as a classical 1D acoustic metamaterial. The radius of the individual scatterer is $a = 0.2 d$, where $d$ indicates the distance between two neighboring cylinders. The filling fraction for this particular cluster is $f = \pi a^2/d^2 \approx 0.1257$.

The waveguide is excited by a monochromatic acoustic monopole $S$ that replaces the center cylinder. The response of the system is obtained numerically by means of a commercial finite element software (Comsol Multiphysics) and using symmetric boundary conditions on the top and bottom edges and perfectly matched layers (PML) on the left and right edges. The frequency of excitation is selected in the first bandgap (see Fig.~\ref{Fig_10} for the general dispersion properties of this waveguide) and has a non-dimensional value $\Omega = 0.0831$. In this excitation regime the diffusive behavior is expected. Remember that, in the absence of disorder or trapping mechanisms and in the range of excitation frequencies where the diffusion approximation holds, the variance of the step-length distribution characterizing the multiple scattering process of the acoustic field is expected to be finite and, if the steps are independent (by virtue of the Central Limit Theorem) the limit distribution should be the Normal distribution as predicted by the standard diffusion model. 

\begin{figure}[h!]
\center{\includegraphics[width=\linewidth]{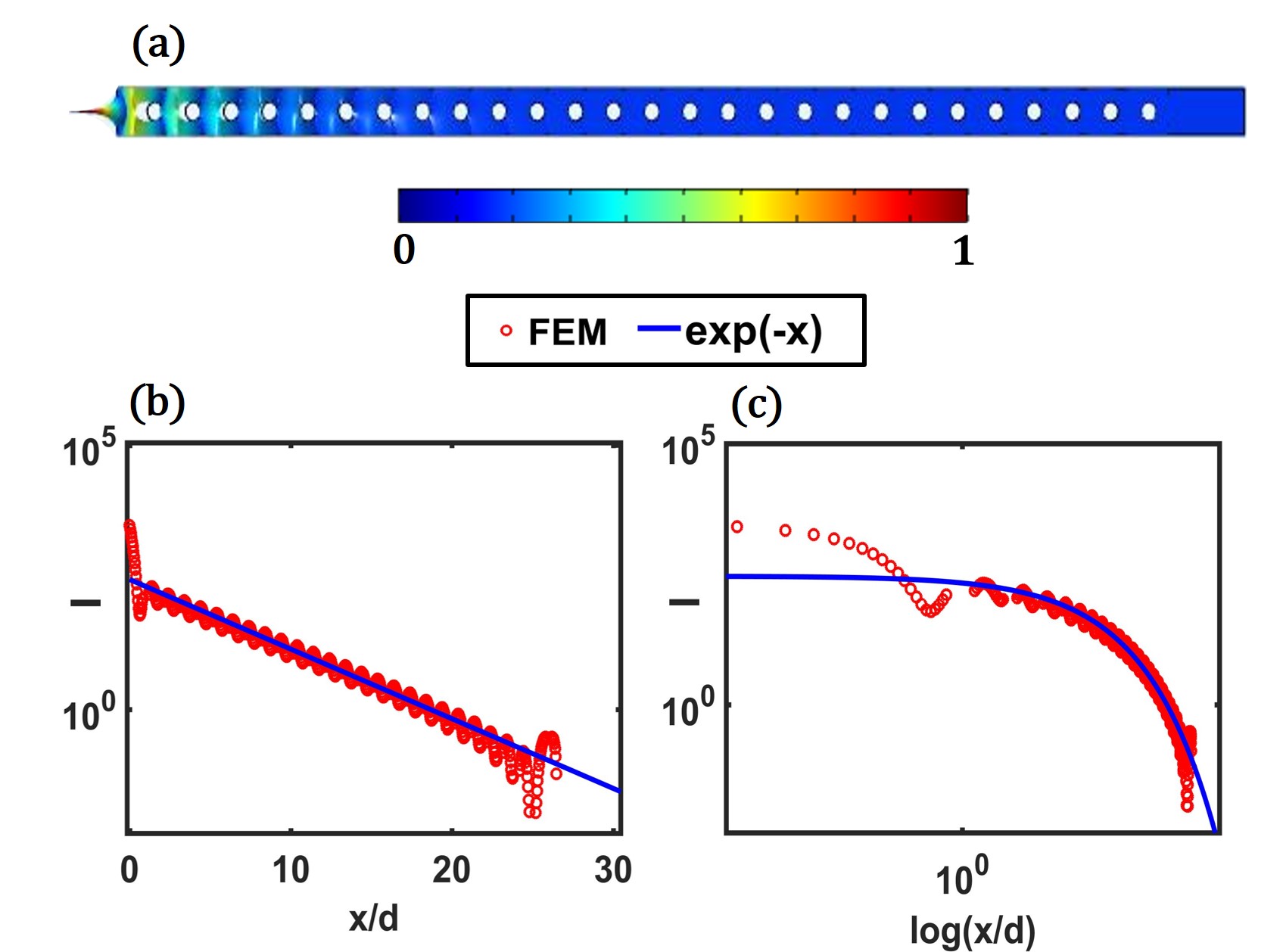}}
  \caption{Numerical predictions of the acoustic intensity in a 1D waveguide with hard scatterers in air subject to a monochromatic monopole excitation. Note that, due to the symmetry of the solution, only the right section of the waveguide is shown. (a) Contour plot of the normalized magnitude of the pressure distribution. (b) Exponential fit of the acoustic intensity in (b) semilogarithmic and (c) logarithmic scales. The trend is well consistent with the expected linear decay predicted by the classical diffusion theory.}
\label{Fig_4}
\end{figure}

The resulting normalized magnitude of the acoustic pressure field generated in the waveguide is shown in Fig.~\ref{Fig_4}(a) in terms of a contour plot and in Fig.~\ref{Fig_4}(b),(c) in terms of the intensity profile along the mid-line of the waveguide as defined later in \S \ref{Modell}. From Fig.~\ref{Fig_4}(b),(c) a characteristic exponential decay of the type $e^{-x/l_s}$ (consistent with the Beer-Lambert law) is very well identifiable. This trend represents the decay of the coherent part of the intensity and corresponds to the squared absolute value of the Green's function solution. In more general terms, this result shows a solution which is perfectly consistent with the classical diffusion behavior. This is a well expected result and it is reported here only for comparison with the results that will be presented below.

\section{Two-dimensional medium}
\subsection{Radial lattice}
\begin{figure}[h!]
\center{\includegraphics[width=\linewidth]{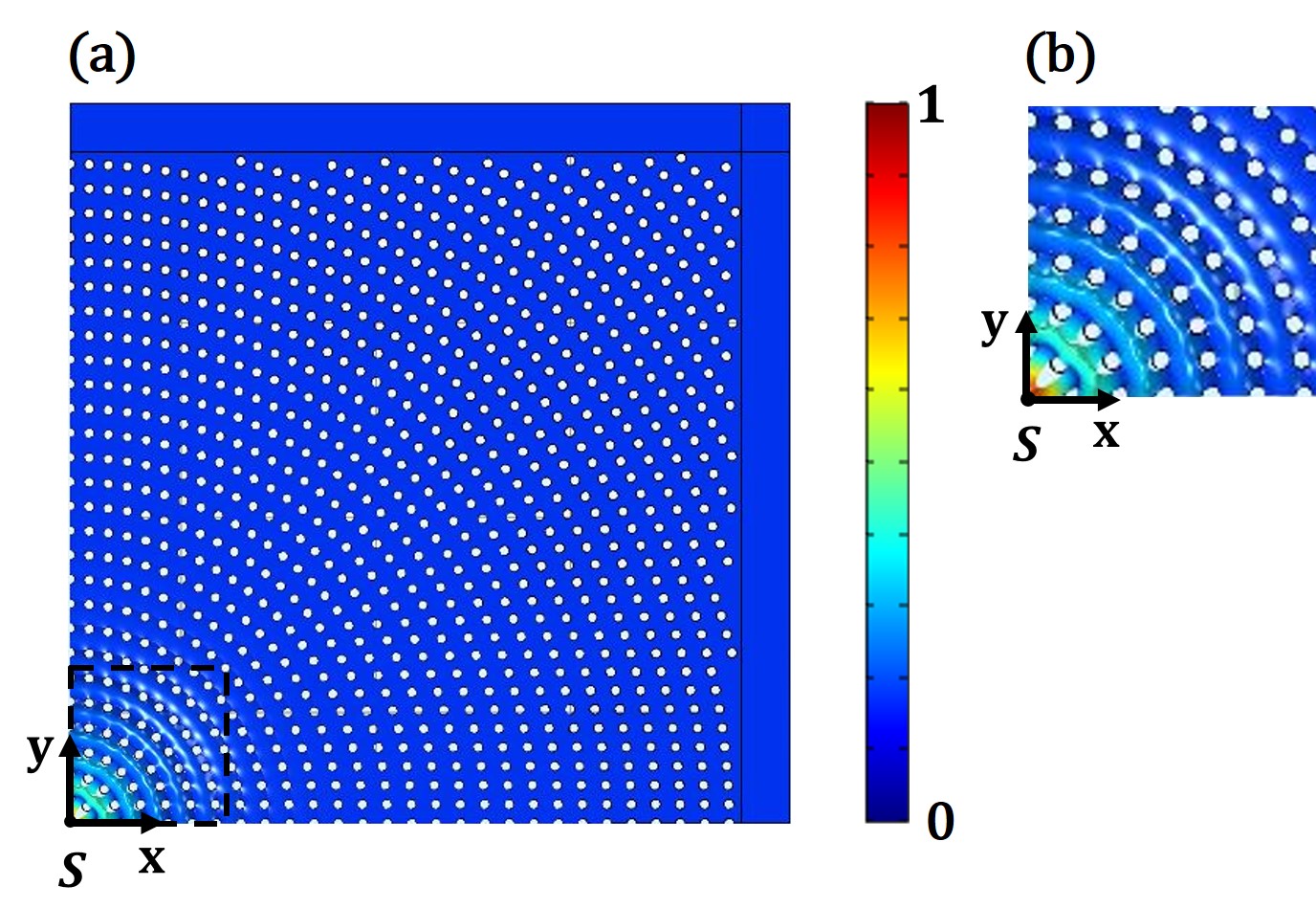}}
  \caption{(a) Top view of the radial lattice and of the corresponding normalized magnitude of the pressure distribution generated by a monopole source located in the center. (b) zoom-in of the field in the dashed black box.}
\label{Fig_6}
\end{figure}

The immediate extension of the previous scenario to a two-dimensional system corresponds to a radial distribution of equally distributed hard scatterers. As in the 1D case, the system is excited by a monochromatic monopole source located in the center of the 2D lattice at point $S$. The source is monochromatic and it is actuated at the non-dimensional frequency $\Omega = 0.0831$, that belongs to the first bandgap.
The normalized magnitude of the acoustic pressure field for this system is numerically calculated and shown in Fig.~\ref{Fig_6}(a). Fig.~\ref{Fig_6}(b) provides a closeup view of the field around the source (the area within the black dashed line).

\begin{figure}[h!]
\center{\includegraphics[width= 6 cm]{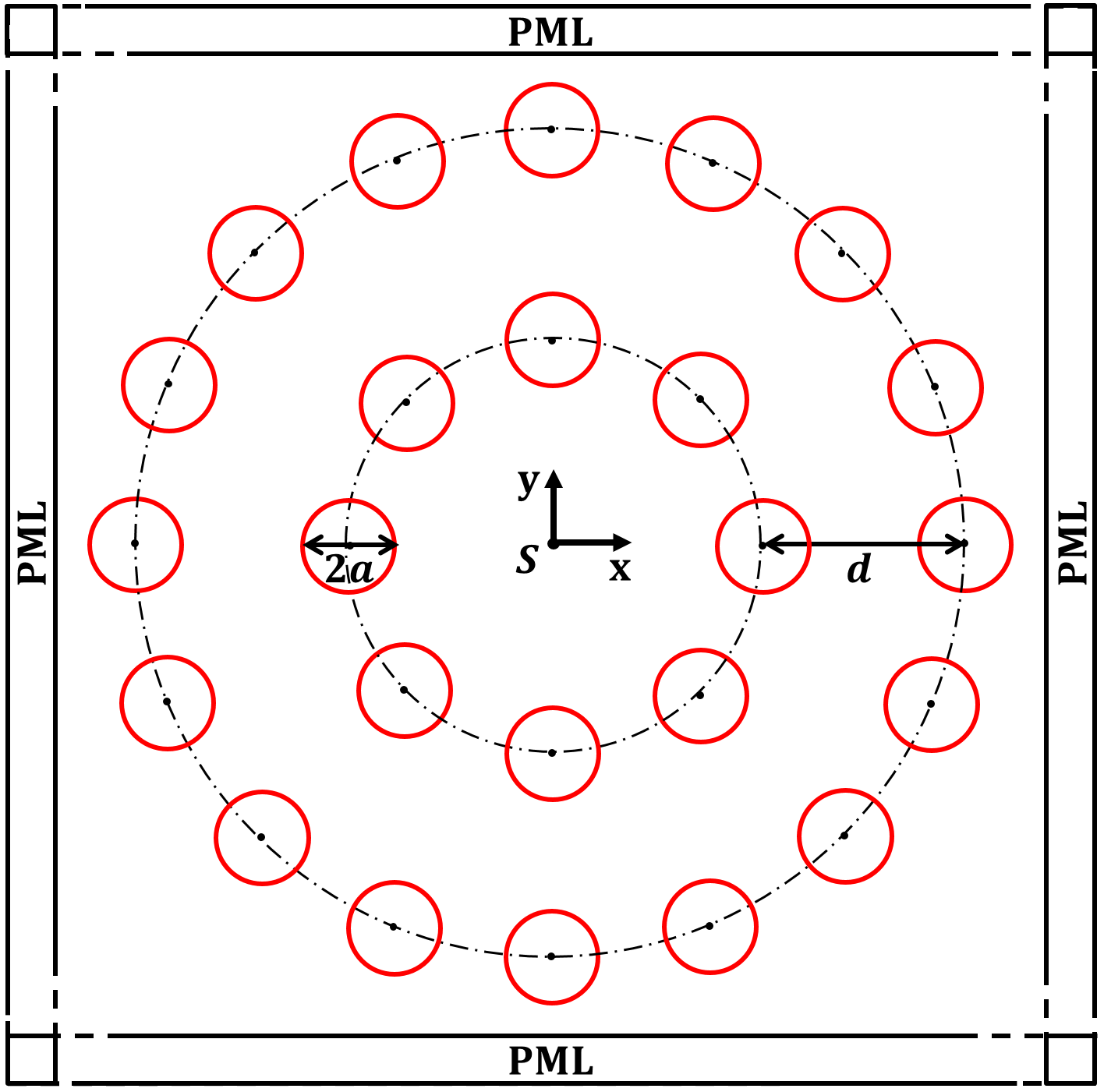}
  \caption{Schematic of the 2D radial distribution of equally spaced hard cylindrical scatterers in air. The medium is excited by a monopole-like acoustic source located at $S$. Perfectly matched layers (PML) are used all around the boundary.}
\label{Fig_5}}
\end{figure}

\begin{figure}[h!]
\center{\includegraphics[width=\linewidth]{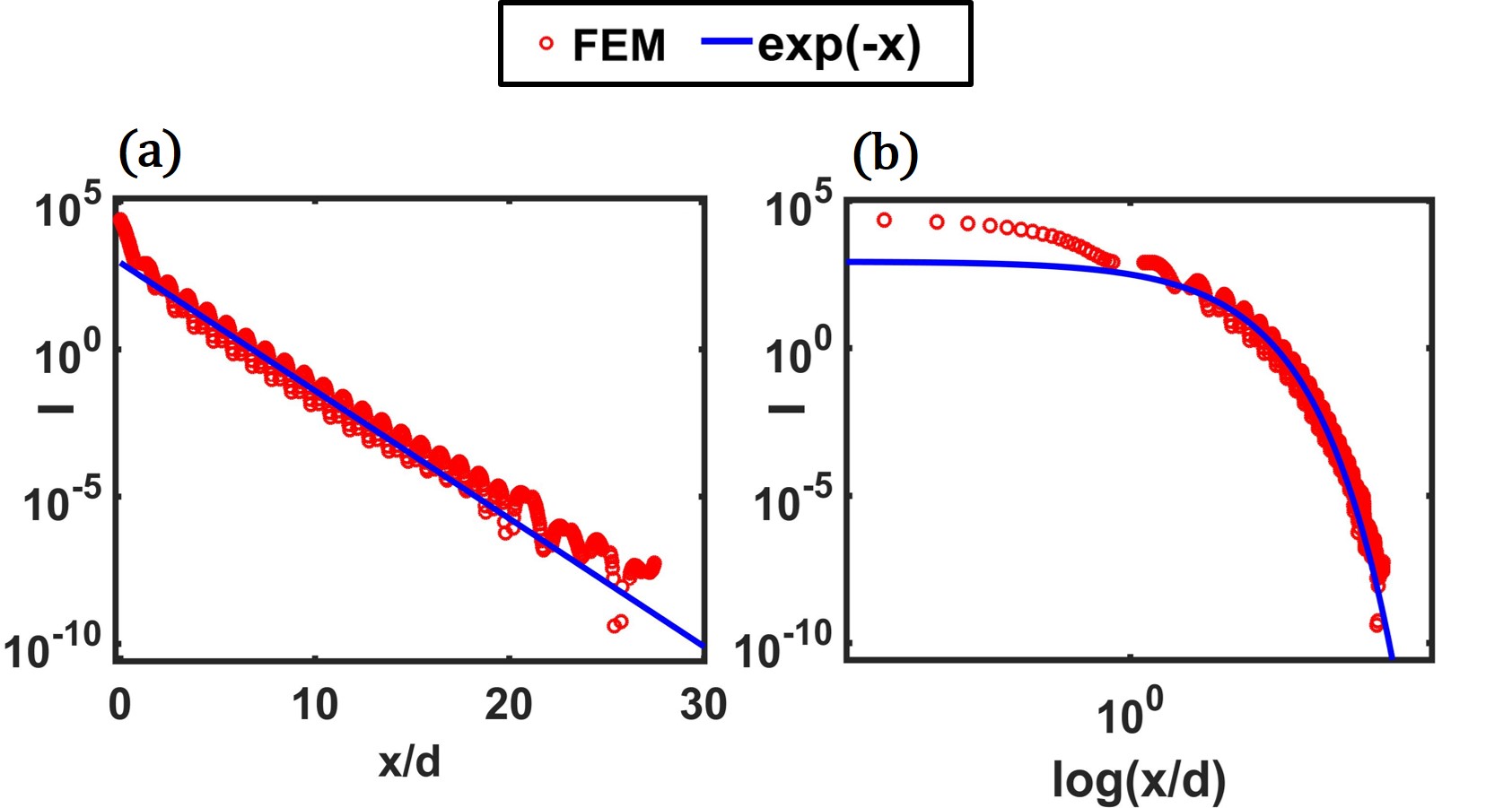}}
  \caption{Exponential fit in (a) semilogarithmic and (b) logarithmic scales of the acoustic intensity of a radial distribution of scatterers excited by a monopole point source.}
\label{Fig_7}
\end{figure}

The acoustic intensity profile along the $x$-axis direction shows an exponential decay as illustrated by Fig.~\ref{Fig_7}. All radial directions (not shown) exhibit an identical response as expected due the azimuthal symmetry of the system. 

As in the 1D case, this linear decay of the intensity distribution was expected and confirms that, in systems with a high degree of symmetry, a classical diffusion behavior should be recovered. From a practical perspective, this radial lattice could be seen as a radial arrangement of 1D waveguides.

\subsection{Rectangular lattice} \label{rect_lattice}

The dynamic behavior of the lattice changes quite drastically when the axial-symmetry is removed. Consider the square lattice of scatterers schematically illustrated in Fig.~\ref{Fig_1}. Assume each scatterer having an individual radius of $a = 0.2 d$, where $d$ is the distance between two neighboring cylinders. The filling fraction for this periodic cluster is $f = \pi a^2/d^2 \approx 0.1257$. 

\subsubsection{Dispersion analysis} \label{DispersionRelation}

\begin{figure}[h!]
	\center{\includegraphics[width=\linewidth]{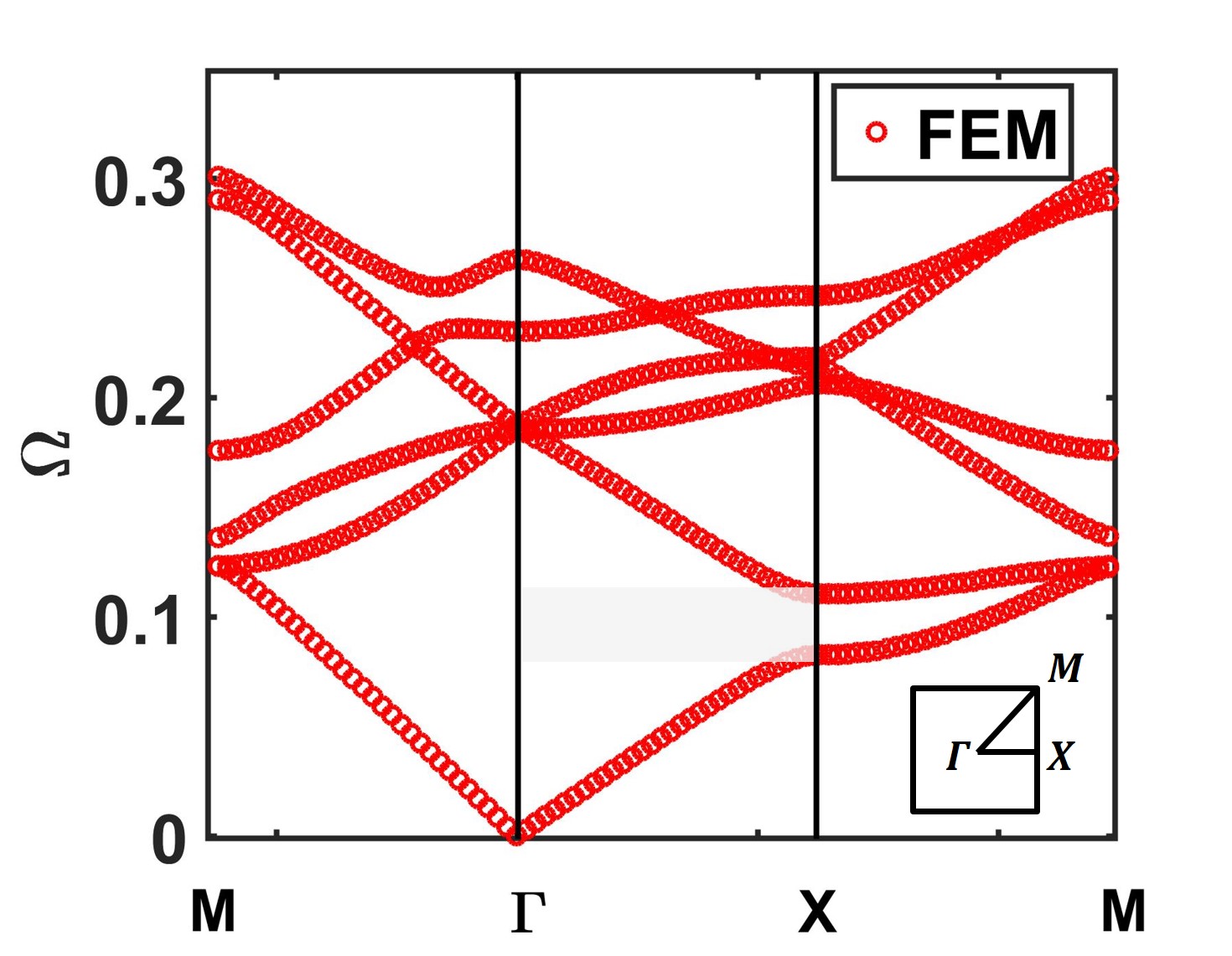}}
	\caption{Dispersion band structure for the square lattice of hard cylinders in air. The gray box indicates the partial bandgap in the $\Gamma-X$ direction.}
	\label{Fig_10}
\end{figure}

In order to understand the dynamic behavior of this lattice and interpret the results that will follow, we start analyzing the fundamental dispersion structure of the square lattice.
The dispersion was calculated using finite element analysis and the band structure is plotted along the irreducible part of the first Brillouin zone, as shown in Fig.~\ref{Fig_10}.

The results highlight the existence of anisotropy in terms of directions of propagation. These directions are connected to the existence of a partial bandgap in the $\Gamma-X$ direction between the non-dimensional frequencies $\Omega = 0.0824$ and $\Omega = 0.1103$. When the system is excited at a frequency within the bandgap, the propagation acquires an anisotropic distribution (see \S~\ref{Forced} and Fig.~\ref{Fig_8}), because propagation can only occur in the $\Gamma-M$ direction. This is not an unexpected result and, in fact, it is fully consistent with the propagation behavior expected in square periodic lattice. However, we will show that these dispersion characteristics play a key role in the occurrence of anomalous behavior.


\subsubsection{Forced response} \label{Forced}
\begin{figure}[h!]
	\center{\includegraphics[width = 9 cm]{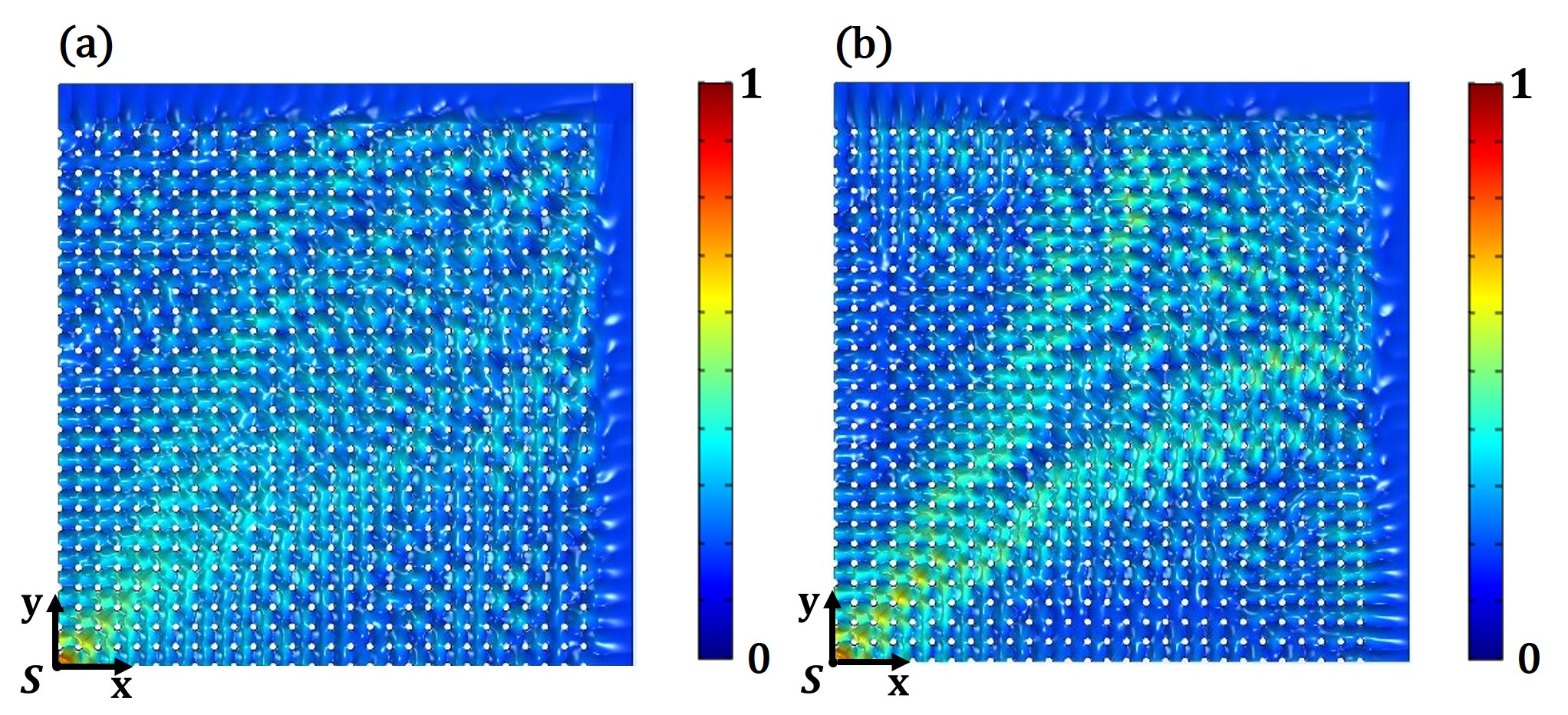}}
	\caption{Normalized magnitude of the pressure distribution corresponding to (a) a  pass-wavelength $\lambda/d = 2.3352$ and (b) to a gap wavelength $\lambda/d = 2.1552$.}
	\label{Fig_8}
\end{figure}

The forced response of the lattice was also numerically evaluated. In this case, the lattice is excited by a monochromatic acoustic monopole $S$ that replaces the center cylinder. As for the previous two lattices, the total acoustic pressure field is calculated numerically using the finite element method and reported in Fig.~\ref{Fig_8}. More specifically, Fig.~\ref{Fig_8}(a) presents the response to an excitation outside the first bandgap, while Fig.~\ref{Fig_8}(b) reports the case just inside the first bandgap. Note that due to symmetry considerations, only a quarter of the domain was solved. 

As the acoustic wave fronts propagate through the medium in the radial directions and interact with the scattering particles, the rays are scattered in multiple directions. In both cases it is evident that the propagation is strongly anisotropic and occurs mostly along the diagonal directions of the lattice.

\begin{figure}[h!]
	\center{\includegraphics[width=\linewidth]{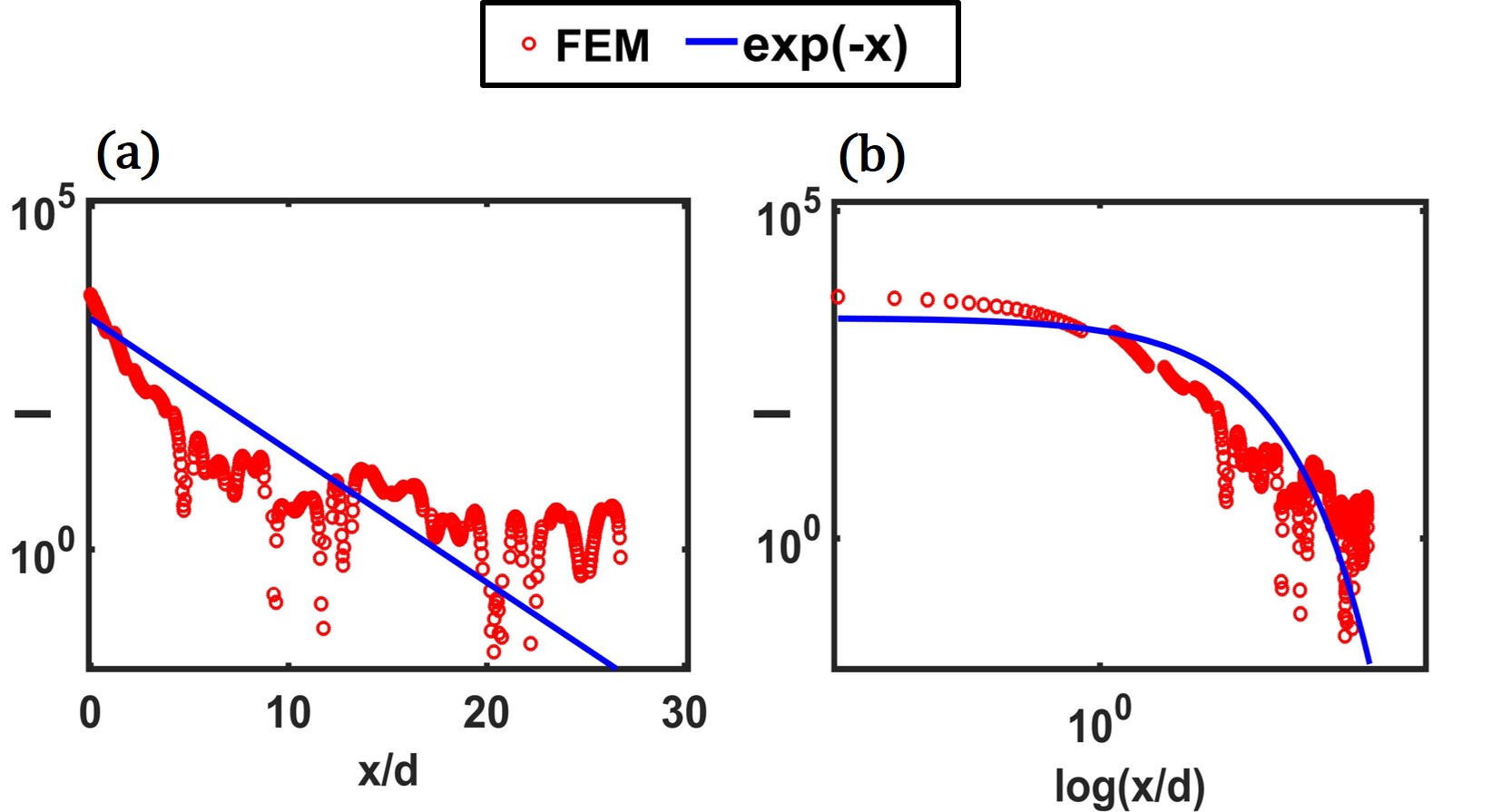}}
	\caption{Exponential fit of the intensity of the pressure field for a periodic lattice of scatterers excited by a monopole point source in (a) semilogarithmic and (b) logarithmic scales.}
	\label{Fig_9}
\end{figure}

The response of the medium is shown in Fig.~\ref{Fig_8} in terms of the normalized magnitude of the acoustic pressure distribution. Contrarily to what observed for the radial lattice, in this case the intensity distribution does not decay linearly. This behavior is very evident by performing a numerical fit of the simulation data, as shown in Fig.~\ref{Fig_9}. These results suggest the occurrence of an unexpected mechanism of diffusion despite the lattice periodicity.

This is a remarkable departure compared with available results in the literature that, to-date, have highlighted the occurrence of anomalous diffusion only in connection with random distributions of geometric or material properties.

\section{Radiative transport approach} \label{Modell}
 The results presented above illustrated that in case of anisotropic propagation a departure from the classical diffusive behavior is observed. In this section, we use a traditional radiative transport approach with renormalization to show that this observed behavior can be mapped to anomalous diffusion.

We investigate the presence of anomalous diffusion for wavelength ranges in the passband and in the bandgap. As already pointed out, within the regime $\lambda/2 \pi<l_{t}<L$ the diffusion approximation applies and the spatial evolution of the wave amplitude can be predicted by a diffusion equation for the wave intensity.

\begin{figure}[h!]
\center{\includegraphics[width = 5 cm]{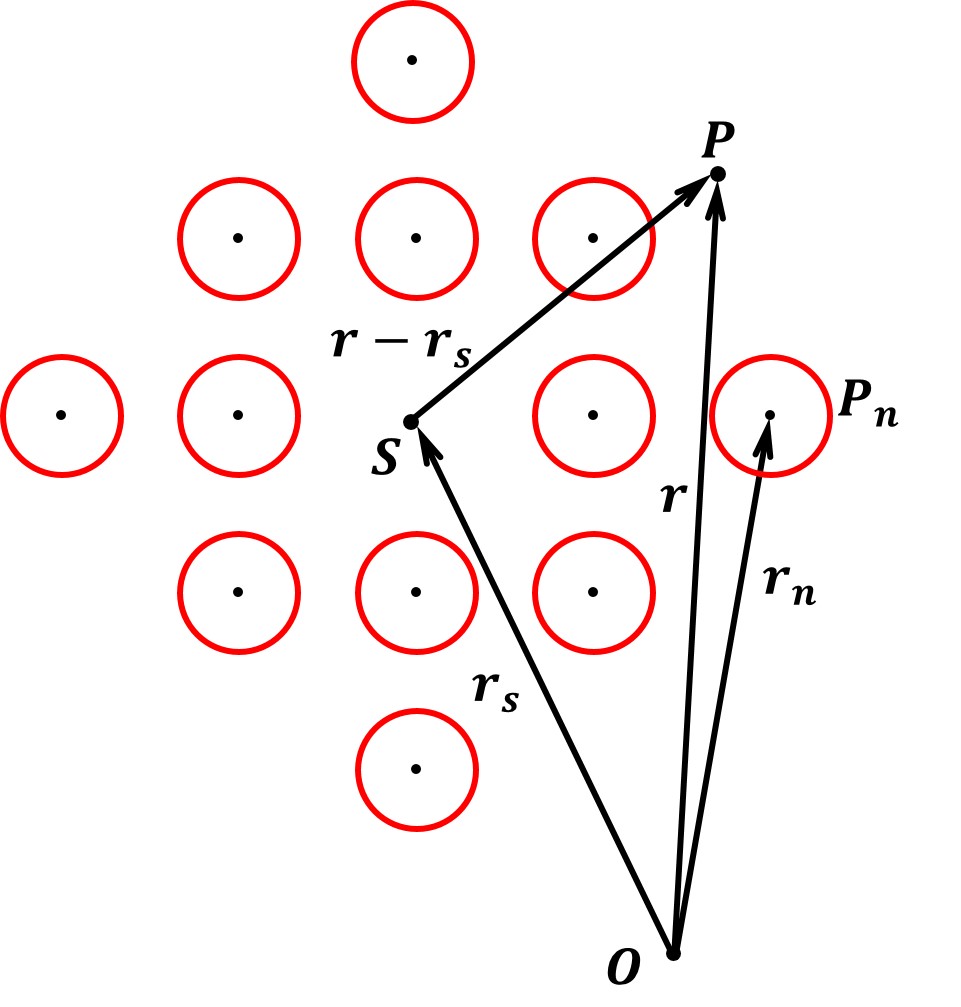}}
  \caption{ Schematic of a cluster of cylindrical hard scatterers used to calculate the radiative transport solution. $\boldsymbol{\vec{r}}_s$ and $\boldsymbol{\vec{r}}$ represent the position vector of the source $S$ and of a generic point $P$, respectively.}
\label{Fig_11}
\end{figure}

Starting from a cluster of particles, as schematically illustrated in Fig.~\ref{Fig_11}, and applying the diffusion approximation the 2D diffusion equation for harmonic excitation and lossless scatterers is given by:

\begin{align} \label{Diffusion equation}
\begin{split}
\nabla^2 I =-\frac{P_0}{\pi l_t}\delta(\vec{r}-\vec{r}_s)
\end{split}
\end{align}

\noindent where $I$ is the intensity of the acoustic wavefield, $P_0$ is the total emitted acoustic power, $\vec{r}$ and $\vec{r}_s$ are the position vectors of the source $S$ and of a generic point $P$, respectively. The average acoustic intensity of a monochromatic monopole source can be obtained as $\langle I \rangle = ||0.5*Re(p \cdot v')||$, where $p$ is the pressure field, and $v'$ is the complex conjugate of the velocity field. The diffusion equation Eq.~(\ref{Diffusion equation}) requires the following boundary conditions at the edge of the domain to be solved:

\begin{eqnarray} 
\begin{split}
I\mathbf{(r_s)}-\frac{\pi l_t}{4}\frac{\partial }{\partial n} I\mathbf{(r_s)} = 0
\end{split}
\end{eqnarray}

\noindent where $\mathbf{\hat{n}}$ is the unit inward normal. These boundary conditions are obtained by the requirement of zero inward flux at the boundaries\cite{ishimaru1978wave}. The numerical value of this boundary condition on the intensity was obtained by the previous finite element model.

By enforcing this boundary condition, Eq.~(\ref{Diffusion equation}) can be solved analytically: 

\begin{align} 
\begin{split}
I = -\frac{P_0}{2\pi^2 l_t}ln\frac{|\vec{r}-\vec{r}_s|}{L}+ I_{0}
\end{split}
\label{Solution}
\end{align}

\noindent where $I_{0}$ is the value of the intensity at the boundary of the cluster of scatterers and $L$ is the size of the computational domain.

\begin{figure}[h!]
	\center{\includegraphics[width=\linewidth]{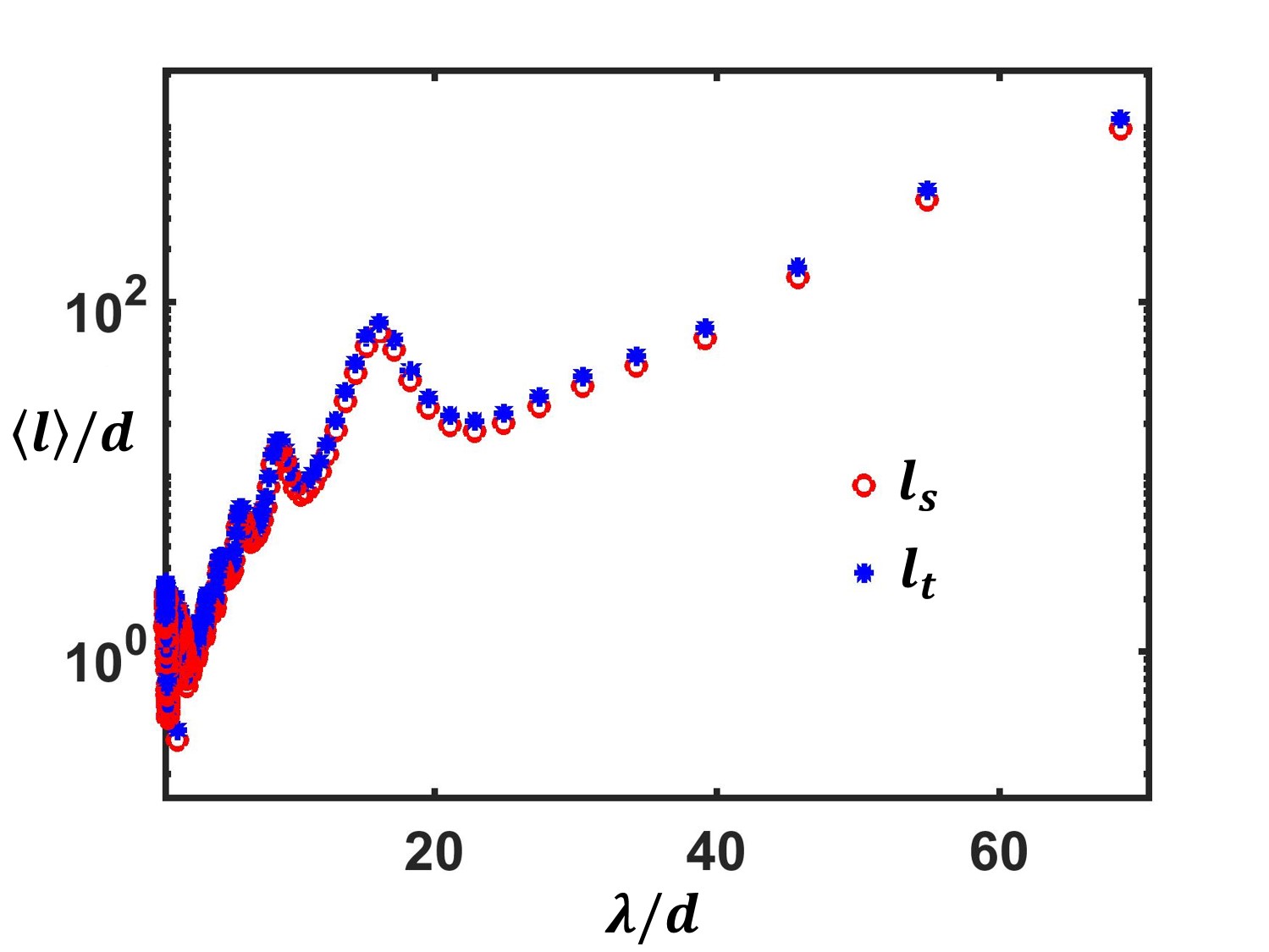}}
	\caption{Non-dimensional scattering mean free path $\langle l_s \rangle$ (red circles) and transport mean free path $\langle l_t \rangle$ (blue stars) versus the non-dimensional wavelength $\lambda/d$. The calculation was performed for a cluster of rigid cylinders of constant radius $a = 0.2 d$ and filling factor $f=0.1257$.}
	\label{Fig_12}
\end{figure}

In order to be able to solve Eq.~(\ref{Solution}), we need to estimate the parameters $l_t$ and $l_s$ and characterize the specific regime of propagation. To achieve this result, we first plot $\langle l_{s}\rangle$ and $\langle l_{t}\rangle$ versus the wavelength $\lambda$ as shown in Fig.~\ref{Fig_12}. These curves were numerically determined using the model presented in \S \ref{rect_lattice} and the Eqs.~(\ref{ls_lt}).

\begin{figure}[h!]
	\center{\includegraphics[width=\linewidth]{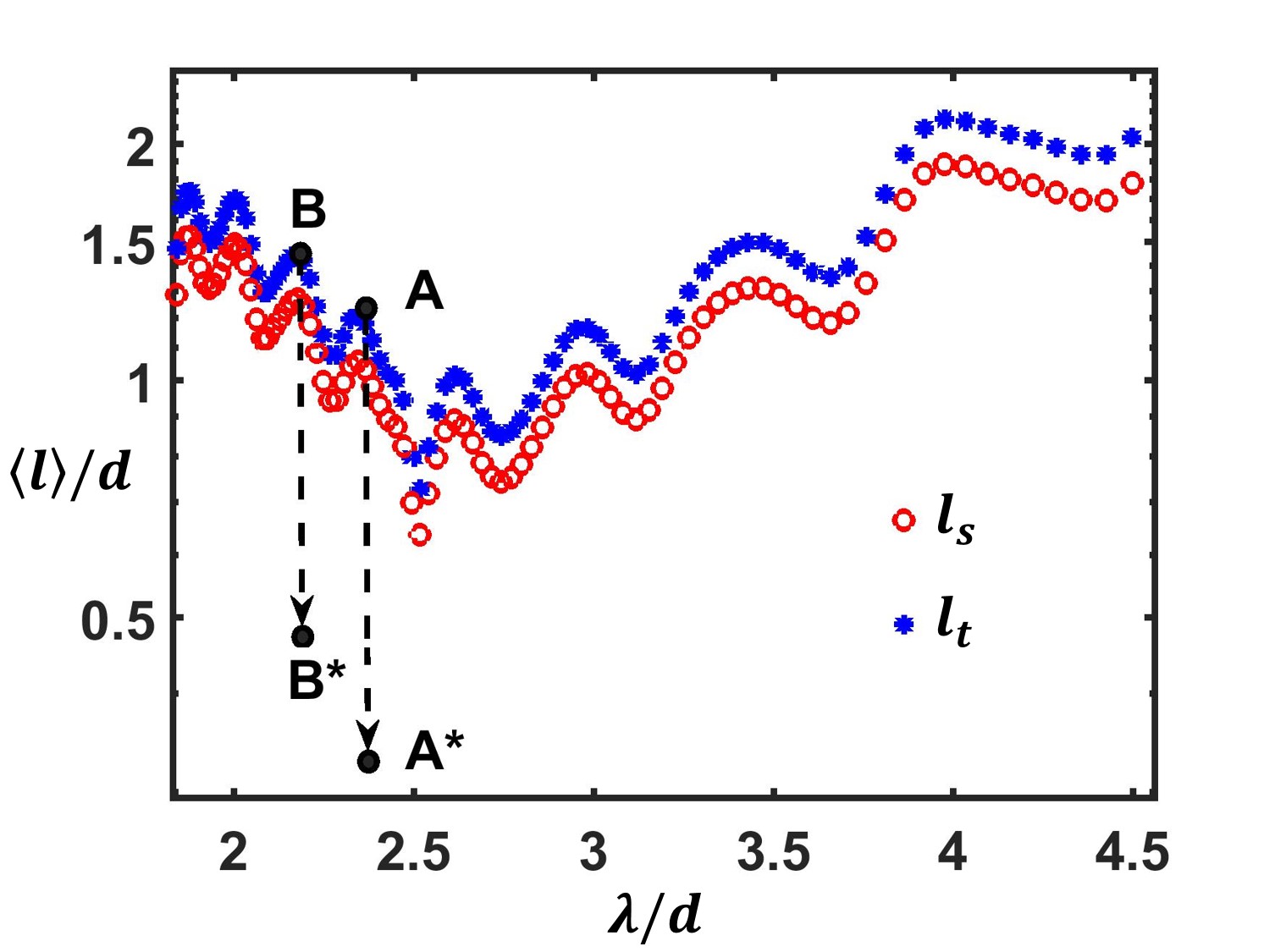}}
	\caption{Detailed view of the transport parameters in the range of frequencies corresponding to the first band gap.}
	\label{Fig_13}
\end{figure}

The transport mean free path $\langle l_{t}\rangle$ is always expected to be greater than $\langle l_{s}\rangle$ and to converge asymptotically to $\langle l_{s}\rangle$ for large wavelengths. In fact, for long wavelengths the wavefield is marginally affected by the presence of the scatterers.
In the short wavelength limit, $l_s/d$ tends to 1 because the wave fronts are highly directional (this is the range of validity of ray acoustics approximation) and \textbf{$l_s$} is approximately given by the average distance between two neighboring scatterers.
Fig.~\ref{Fig_13} shows a detailed view of the previous curves in the frequency range corresponding to the first bandgap and within the diffusive regime. The labels $A$ and $B$ indicate the non-dimensional wavelengths corresponding to the excitation conditions analyzed in the following sections.
Note that these curves provide the foundation to investigate the different regimes of propagation and to implement the renormalization approach.

\subsubsection{Renormalization and anomalous diffusion}  \label{Numerical_1}

\begin{figure}[h!]
\center{\includegraphics[width=\linewidth]{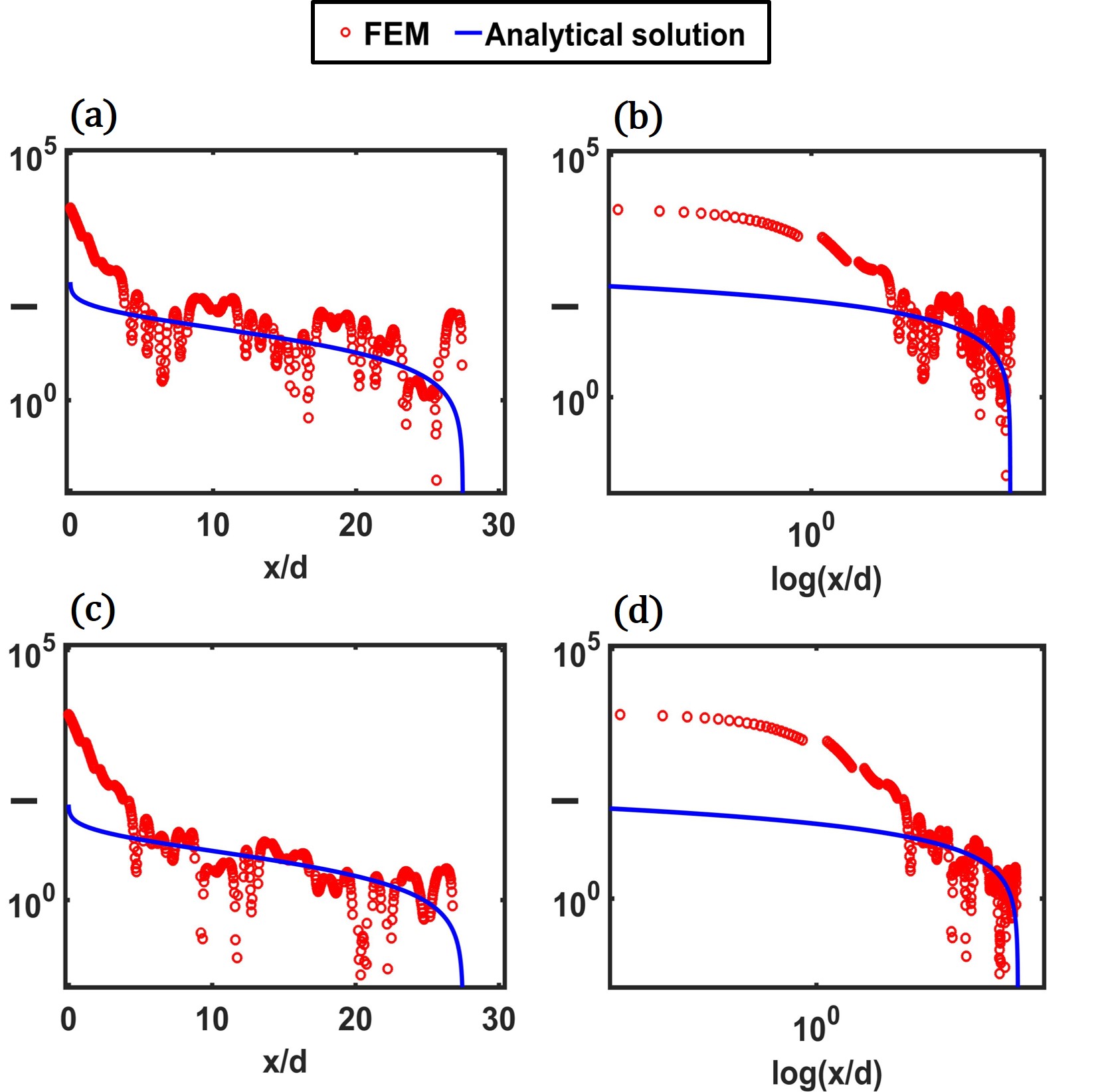}}
  \caption{Acoustic intensity versus distance for a monopole point source at the center of the periodic lattice. (a) semilogarithmic and (b) logarithmic scale plot of $I$ corresponding to a non-dimensional wavelength $\lambda/d = 2.3352$ belonging to the passband. The transport mean free path renormalizes from the value $\langle l_{t}\rangle/d\approx 1.211$ to the value $\langle l_{t}\rangle/d\approx 0.32$. (c) semilogarithmic and (d) logarithmic scale plot of $I$ for a non-dimensional wavelength $\lambda/d = 2.1552$ just inside the bandgap. The transport mean free path renormalizes from the value $\langle l_{t}\rangle/d\approx 1.42$ to the value $\langle l_{t}\rangle/d\approx 0.48$. In both cases the system is in the anomalous diffusion regime.}
\label{Fig_14}
\end{figure}

Fig.~\ref{Fig_14} shows the acoustic intensity distribution $I$ along the $x$ axis for the two excitation conditions identified by the labels A and B.

The red circles show the numerical solution obtained by the FE model and provide a one-dimensional section of the data in Fig.~\ref{Fig_8} along the $x$-axis. The continuous blue line is the analytical solution of the diffusion equation Eq.~(\ref{Solution}) after having rescaled the transport mean free path. In particular, for excitation wavelengths in the first passband the value $\langle l_{t}\rangle/d\approx 1.211$ was rescaled to $\langle l_{t}\rangle/d\approx 0.32 \pm 0.02$ (label $A^*$ in Fig.~\ref{Fig_13}), while for the first bandgap the value $\langle l_{t}\rangle/d\approx 1.42$ was rescaled to $\langle l_{t}\rangle/d\approx 0.48 \pm 0.02$ (label $B^*$ in Fig.~ \ref{Fig_13}). 

These results show that, in order to be able to predict the numerical data by using the diffusion approximation, a renormalization of the transport mean free path (and consequently of the diffusion coefficient) must take place. The renormalization requires smaller values of the transport parameters which is a clear indication of superdiffusive anomalous behavior.

\section{Causes of anomalous diffusion}  \label{NumericalViewPer}

In the previous sections we showed the occurrence of anomalous diffusion of acoustic waves in perfectly periodic square lattices and suggested that the possible origin of this mechanism is linked to the anisotropy of the dispersion properties (i.e. to the anisotropy of the bandgaps).

In this section we will present theoretical and numerical models with the intent of uncovering the physical mechanism leading to this unexpected propagation modality. It is anticipated that the occurrence of anomalous diffusion will be connected to the existence of long range interactions between different directions of propagation governed by either bandpass or stopband behavior. We will use a combination of both deterministic and stochastic methods in order to quantify the long-range interactions and to demostrate that they are at the origin of the macroscopic anomalous diffusion mechanism.

More specifically, we will use a scattering matrix approach to quantify the interaction between different scatterers in different regimes. Then, we will use a discrete random walk diffusion model (which uses probability density functions obtained from the scattering model) to show that, under these assumptions, the anomalous diffusion process matches well with the numerically predicted behavior.

\subsection{The scattering matrix}
In order to evaluate and quantify the strength of the interaction between different scatterers in the lattice, we use a multiple scattering approach based on the multipole expansion method. According to this method, after applying the Jacobi's expansion and the Graf's addition theorem, the general solution of the wave field can be expressed as:

\begin{eqnarray}
\begin{split}
\label{Eq.206}
p(\vec{r}_m)=\sum_{j=-\infty}^{\infty} (e^{i\vec{k}\cdot \vec{P}_m} e^{ij(\pi /2-\psi_0)}J_j(\vec{r}_m)+A_j^m H_j(\vec{r}_m))+\\
\sum_{n=1,n\neq m}^{N}\sum_{q=-\infty}^{\infty}A_n^q\sum_{j=-\infty}^{\infty}H_{q-j}(kR_{nm})e^{i(q-j)\Phi_{nm}}J_j(\vec{r}_m)
\end{split}
\end{eqnarray}
where $\vec{k}$ represents the wave vector, $\psi_0$ is the angle of the impinging wave field with respect to the $x$-axis, $\vec{P}_m$ is the position vector of the  scatterer's center $O_m$ with respect to the origin $O$ of the system of reference, $\vec{r}_m$ is the position vector of a generic point $P$ with respect to the scatterer's center $O_m$, $R_{nm} = \left |\vec{P}_m - \vec{P}_n \right |$, $J_q(\cdot)$ are the Bessel functions of the first kind, $H_q(\cdot)$ are the Hankel functions of the first kind.
 
\begin{figure}[h!]
	\center{\includegraphics[width = 5 cm]{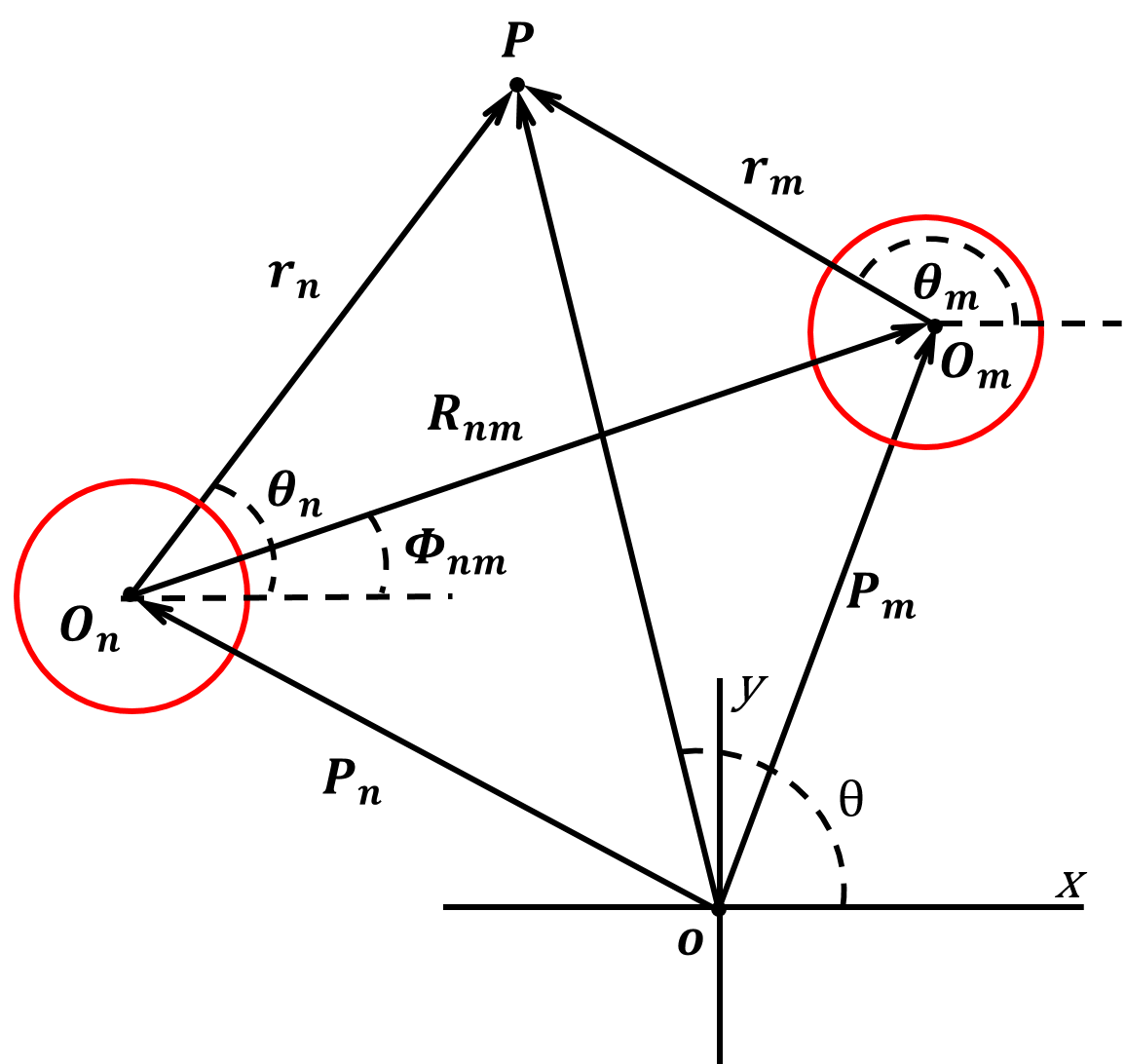}}
	\caption{Schematic of the generic $n^{th}$ and $m^{th}$ cylinders used in the multipole expansions method for multiple scattering}
	\label{Fig_15}
\end{figure}

To determine the unknown amplitude coefficients $A_m^q$ the boundary conditions at the surface of the $m$th cylinder must be enforced. The result is a linear set of equations as follows\cite{linton2005multiple,martin2006multiple,kafesaki1999multiple}:

\begin{eqnarray}
\begin{split}
\label{Eq.207}
A_m^p+Z_p\sum_{n=1,n\neq m}^{N}\sum_{q=-\infty}^{\infty}A_n^qH_{q-p}(kR_{nm})e^{i(q-p)\Phi_{nm}}= \\ -Z_p e^{i\vec{k}\cdot\vec{P}_m} e^{ip(\pi/2-\psi_0)},\\ \quad m=1,...,N ,\quad p=0,+1,-1,...
\end{split}
\end{eqnarray}

\noindent where $Z_p =J'_p(\cdot)/ H'_p(\cdot)$ specifies the Neumann boundary conditions on the surface of the cylinders. The unknown amplitude coefficients $A_m^q$ can be determined by solving the infinite system of algebraic equations with inner sum truncated at some positive integer $|q| =Q$. The information about the relative energy exchange between the scatterers can be obtained by rearranging Eq.~(\ref{Eq.207}) in matrix form as: 

\begin{eqnarray}
\begin{split}
\label{Eq.209}
(I-TS)f=Ta
\end{split}
\end{eqnarray}

\noindent where $I$ is the unit matrix, $T$ is the block diagonal impedance matrix, the vector $f$ represents the unknown expansions of scattered waves, and the vector $a$ stands for the expansion vector of incident waves on all the scattering cylinders. Finally the matrix $S$ is the so called combined translation matrix that can be expressed as follows:

\begin{eqnarray}
\begin{split}
\label{Eq.210}
S=\begin{bmatrix}
0 & L_{12} &...& L_{1N}\\ 
L_{21} & 0 &...&L_{2N}\\ 
\vdots  & \vdots  & \ddots  & \vdots \\ 
L_{N1}& L_{N2} & ... & 0
\end{bmatrix}
\end{split}
\end{eqnarray}

\noindent where the matrix $L_{nm}$ is defined as follows:

\begin{eqnarray}
\begin{split}
\label{Eq.211}
L_{nm}(q,p)=H_{q-p}(kR_{nm})e^{i(q-p)\Phi_{nm}}.
\end{split}
\end{eqnarray}

The matrix $L_{nm}$ represents the translation matrix between the $n$th and the $m$th cylinder, representing therefore the incident wave on the $n$th cylinder caused by the scattered wave off the $m$th cylinder. The elements of the translation matrix can be obtained from the addition theorem of cylindrical harmonics also known as Graf's theorem.

The generic term $S_{mn}$ quantifies the portion of the acoustic intensity scattered by the cylinder $m$ capable of reaching the cylinder $n$. Equivalently, it represents the fraction of the acoustic intensity reaching the cylinder $n$ due to the wave scattered by the cylinder $m$.

\begin{figure}[h!]
\center{\includegraphics[width= 7 cm]{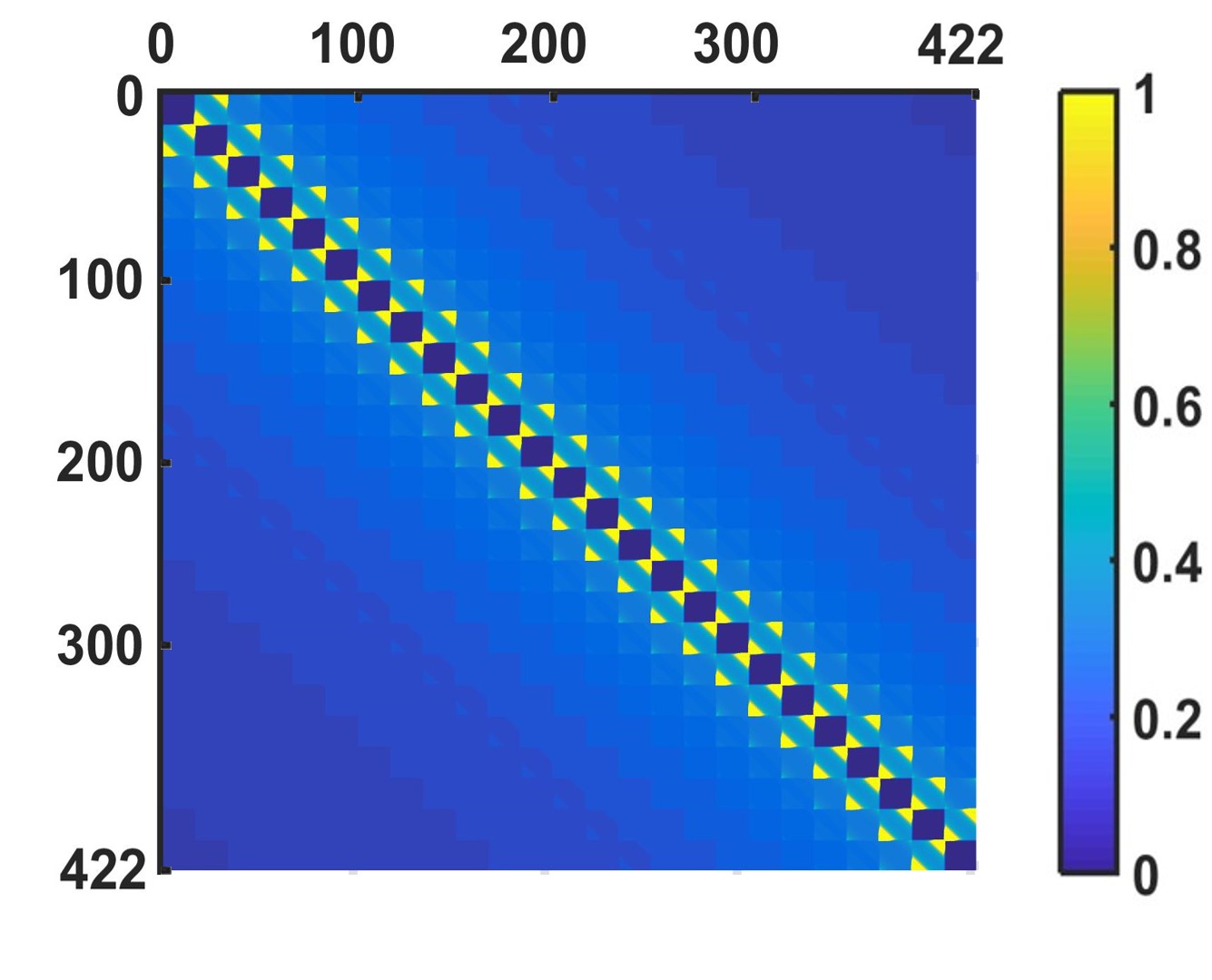}}
  \caption{Scattering matrix for the 1D waveguide. Results correspond to a filling factor of $0.1257$ and a non-dimensional excitation wavelength $\lambda/d =2.1552$. The scattering cylinders have a constant radius equal to $a = 0.2 d$, where $d$ indicates the distance between two neighboring cylinders.}
\label{Fig_16}
\end{figure}

This approach was applied to model both the 1D and the 2D waveguides. The normalized scattering coefficients for the 1D waveguide are shown in matrix form in Fig.~\ref{Fig_16}. Each block of this matrix has a size $\bar{Q} = 2*Q+1$, where $Q$ is the total number of spherical harmonics used in the multipole series expansion. The total size of the matrix is $N*\bar{Q}$, where $N$ is the total number of cylinders in the cluster. The main diagonal represents the coefficient $S_{mm}$, that is the scattering of a given cylinder $m$ towards itself, and therefore these terms are all zero.

The analysis of Fig.~\ref{Fig_16} shows, as expected, that in a 1D waveguide the scatterers only interact with their closest neighbors. In other terms, there is no evidence of long-range interaction in 1D periodic waveguides. This is not surprising because we had already found from the full-field simulations in Fig.~\ref{Fig_4}(a) that the diffusive transport was following a purely Gaussian distribution (hence dominated by nearest neighbor interactions). 

\begin{figure}[h!]
\center{\includegraphics[width= 7 cm ]{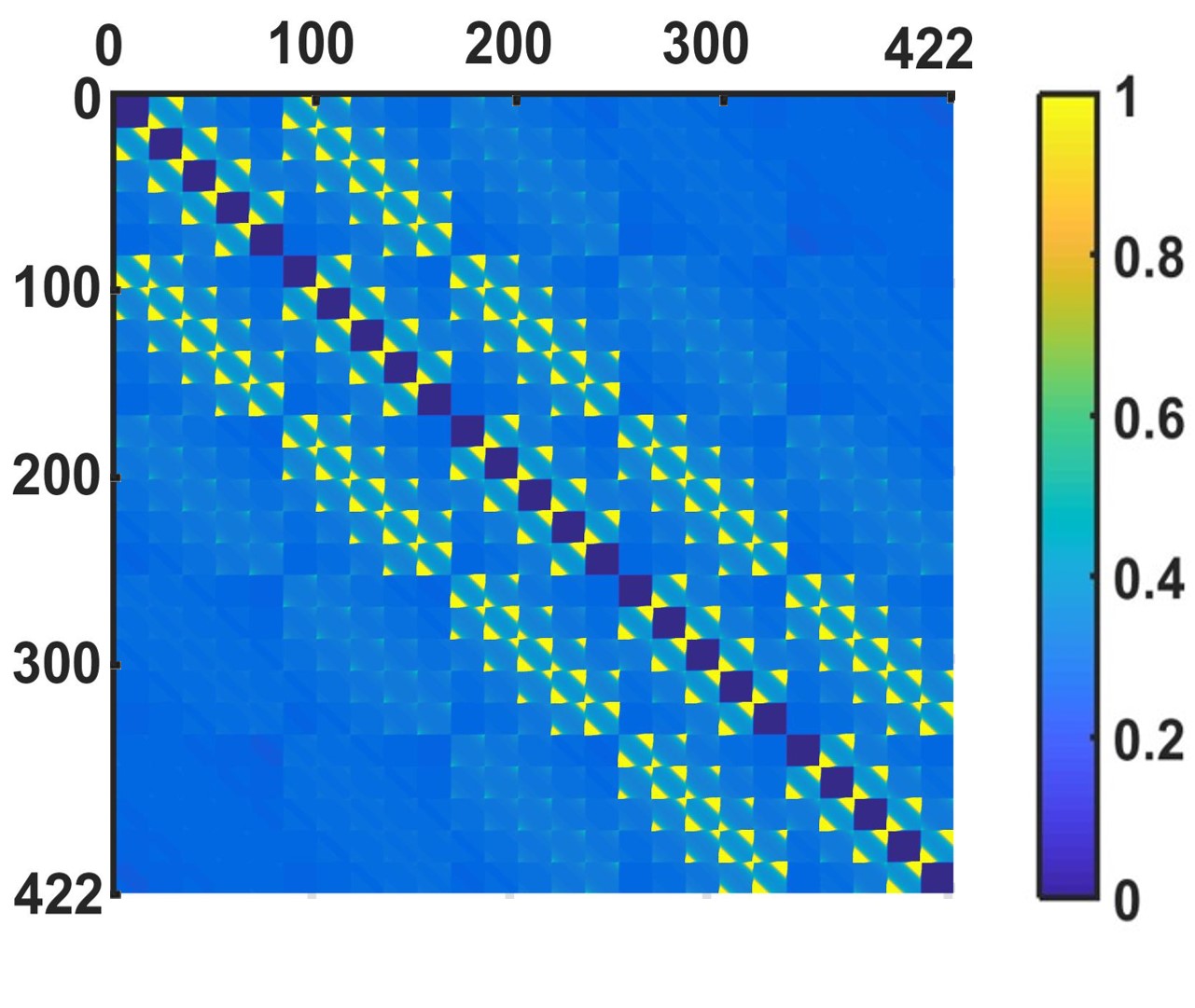}}
  \caption{Scattering matrix for the 2D waveguide made of a square lattice of 25 scatterers. Results corresponding to a filling factor of $0.1257$ and a wavelength $\lambda/d = 2.1552$.}
\label{Fig_17}
\end{figure}

In a similar way, the analysis can be repeated for the 2D waveguide with square lattice structure. The resulting scattering coefficients are shown in 
Fig.~\ref{Fig_17}. Contrarily to the 1D example, this scattering matrix has the appearance of a tridiagonal matrix that highlights the substantial interactions between distant neighbors. In other terms, the rectangular lattice show strong evidence of long-range interactions.
These results provide a first important observation concerning the cause of anomalous diffusion in periodic rectangular lattice, that is the anisotropy of the dispersion bands gives rise to long-range interactions that ultimately alter the diffusion process. 

\subsection{Discrete random walk models: approximate acoustic intensity}

The previous analysis is not yet sufficient to provide conclusive evidence that the long-range interactions due to the bandgap anisotropy are the main cause of the anomalous wave diffusion. In order to identify this further logical link, we developed a discrete random walk (DRW) model capable of simulating the diffusion process resulting from the multiple scattering of the acoustic waves. 

The interaction between the different elements of a DRW model is typically represented by probability density functions (\textit{pdf}). In the following, the \textit{pdf}s are synthesized based on the coefficients of the scattering matrix. The model can then be numerically solved in order to predict the approximate acoustic intensity resulting from the scattered field.

\subsubsection{1D discrete random walk model}
\begin{figure}[h!]
	\center{\includegraphics[width=\linewidth]{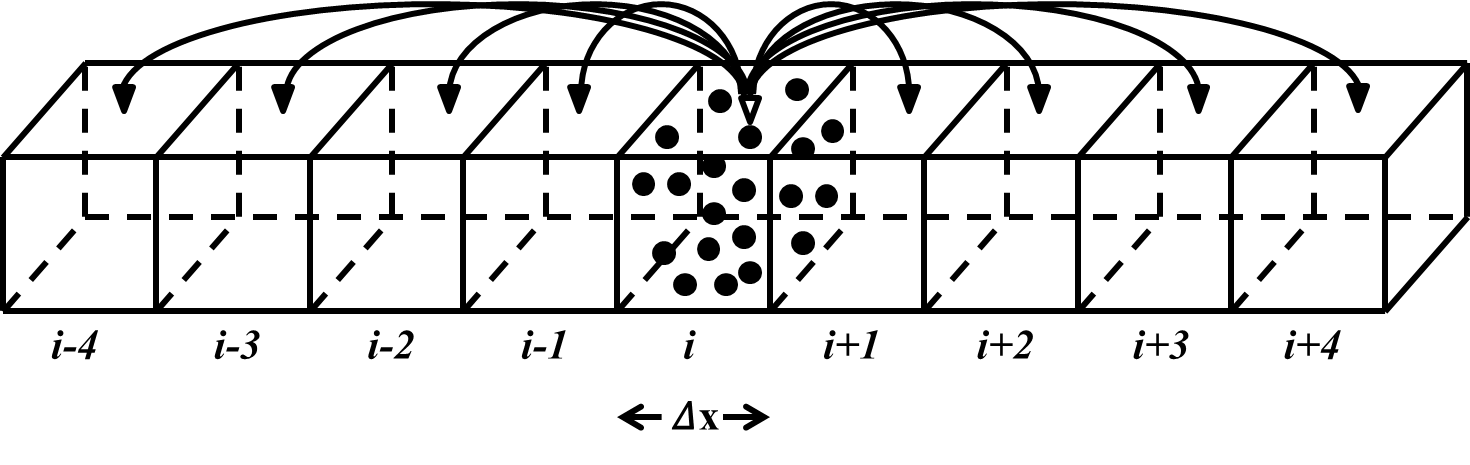}}
	\caption{Schematic of the 1D discrete random walk (DRW) model used to describe the anomalous diffusion of acoustic rays in a scattering medium. This is often called a "box model". Every box represents a scatterer, while the dots represent the impinging acoustic rays that are scattered towards different boxes according to the selected \textit{pdf}.}
	\label{Fig_18}
\end{figure}

The DRW model for a 1D waveguide is composed of a series of boxes (see Fig.~\ref{Fig_18}), each one representing a scatterer. This model can be seen as the direct discrete equivalent of the 1D waveguide in Fig.~\ref{Fig_3}. The dots in each box represent the different acoustic rays impinging on a given scatterer and being refracted towards different (scattering) elements. This model follows a ray acoustic approximation which is a reasonable assumption in the range of wavelength we have been considering.
To simulate the monopole acoustic source located at the center of the waveguide, the center box (labeled $i$) contains a source term that serves as an omni-directional source of rays. In the 1D model, the rays emitted from the center box can be scattered both to the left and to the right according to the associated \textit{pdf} synthesized based on the elements of the scattering matrix.

At every time increment, the rays "jump" into another box following a Markovian process and a \textit{pdf} proportional to the coefficients extracted from the scattering matrix. The equilibrium condition needed to solve the DRW model and simulate the evolution of the acoustic intensity upon scattering is given by imposing the conservation of rays:

\begin{align} \label{eqn:Conservation number of particles}
\begin{split}
n_{i,j+1}=n_{i,j}+\sum_{k=1}^{N_L} n_{k,j}P(i-k)+\sum_{k=1}^{N_R}n_{k,j}P(k-i)- \\
\sum_{k=1}^{N_L}n_{i,j}P(i-k)-\sum_{k=1}^{N_R}n_{i,j}P(k-i) + B_i
\end{split}
\end{align}

\noindent where $i$ is the box index, $j$ is the time index, $n(i,j)$ is the number of rays at time $i$ entering the box $j$ (i.e. impinging on the scatterer $j$), $B_i$ is the source term, and $N_L$ and $N_R$ represent the number of boxes on the left and right side, respectively. 

The previous equation can be rearranged as follows: 
\begin{align} \label{eqn:Conservation number of particles 2}
\begin{split}
n_{i,j+1}=n_{i,j}+\sum_{k=1}^{N_L}(n_{k,j}-n_{i,j})P(i-k)- \\
\sum_{k=1}^{N_R}(n_{k,j}-n_{i,j})P(k-i)+ B_i.
\end{split}
\end{align}

\begin{figure}[h!]
\center{\includegraphics[width=\linewidth]{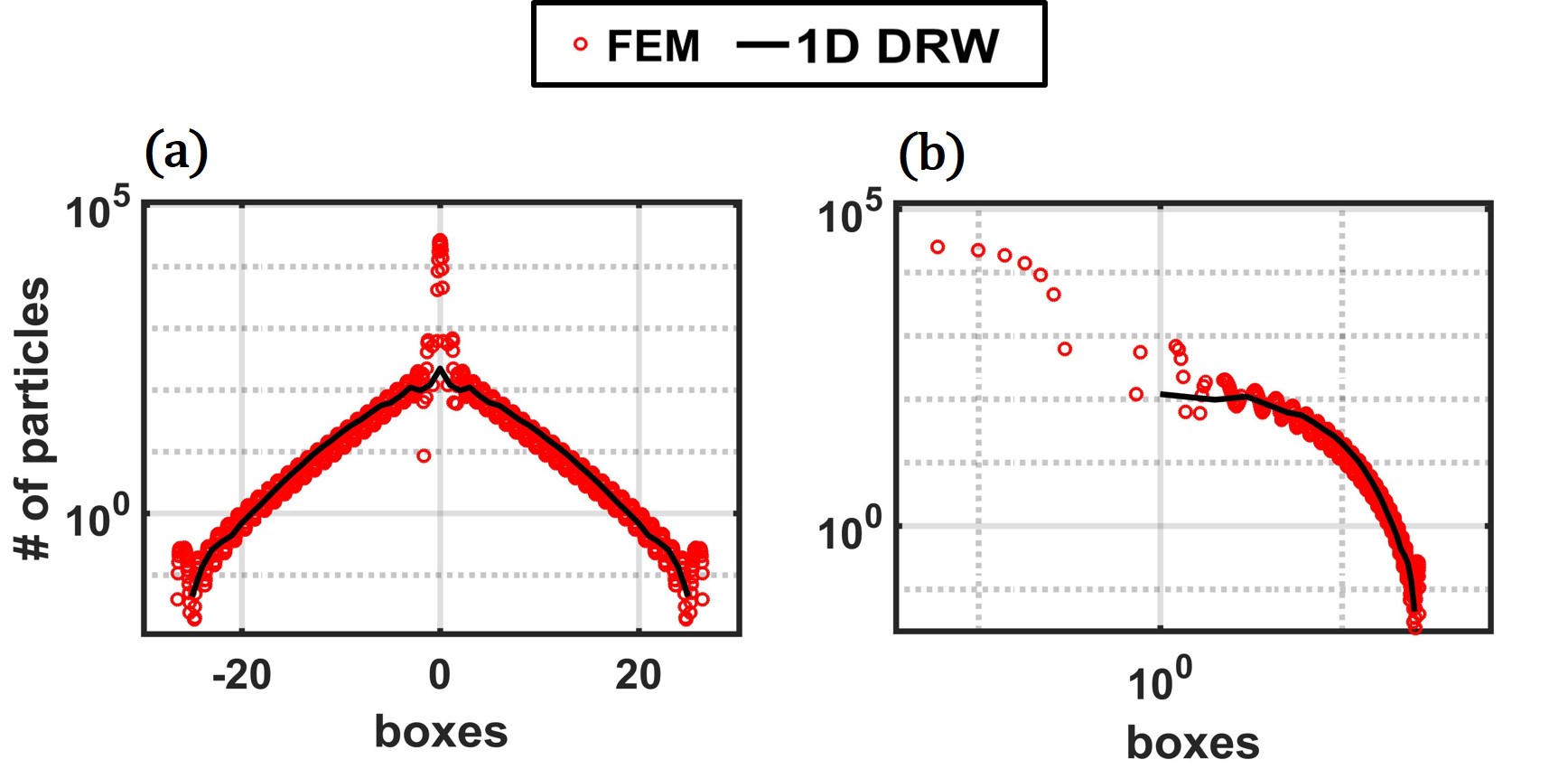}}
  \caption{Comparison of the intensity distribution of a$1$-D waveguide of scatterers with a monopole acoustic source at the center obtained with FEM and the equivalent $1$-D box model in (b) semilogarithmic and (c) logarithmic scales.}
\label{Fig_19}
\end{figure}

The comparison between the intensity distributions obtained with the FE model and by the equivalent 1D DRW model is given in Fig.~\ref{Fig_19}.  

The direct comparison of the results shows a very good agreement between the two models. Note that the DRW is a diffusive model therefore the comparison between the intensity distributions is meaningful only in the tail region. As expected the tails evolve according to a Gaussian distribution. 
The comparison with the 1D waveguide was provided to illustrate the validity of the proposed approach and to confirm that, under the given assumptions, the results from the DRW converge to the full-field simulations.

\subsubsection{2D discrete random walk model}

The same approach illustrated above for the 1D waveguide can be applied to the analysis of the 2D square lattice.
In this case, the DRW model is composed of a 2D distribution of boxes simulating the scatterers. The interactions between different boxes are again expressed in terms of \textit{pdf}s that are synthesized based on the scattering coefficients obtained from the 2D multipole expansion model (Fig.~\ref{Fig_17}).
The equilibrium condition for the 2D DRW model is given by:

\begin{align} \label{eqn:Conservation number of particles 2D}
\begin{split}
n_{i,h,j+1}=n_{i,h,j}+ \\ \sum_{k_h=1}^{N_D}\sum_{k_i=1}^{N_L}(n_{k_i,k_h,j}-n_{i,h,j})P(i-k_i,h-k_h)-\\ \sum_{k_h=1}^{N_U}\sum_{k_i=1}^{N_R}(n_{k_i,k_h,j}-n_{i,h,j})P(k_i-i,k_h-h)+ B_{ih}
\end{split}
\end{align}

\noindent where $i$ and $h$ are the box indices, $j$ is the time index, $n(i,h,j)$ is the number of particles at time $j$ in the box $(i,h)$, $B_{ih}$ is the source term and $N_L $, $N_R$, $N_U$ and $N_D$ represent the number of boxes on the left, right, up and down sides, respectively. 

\begin{figure}[h!]
\center{\includegraphics[width=\linewidth]{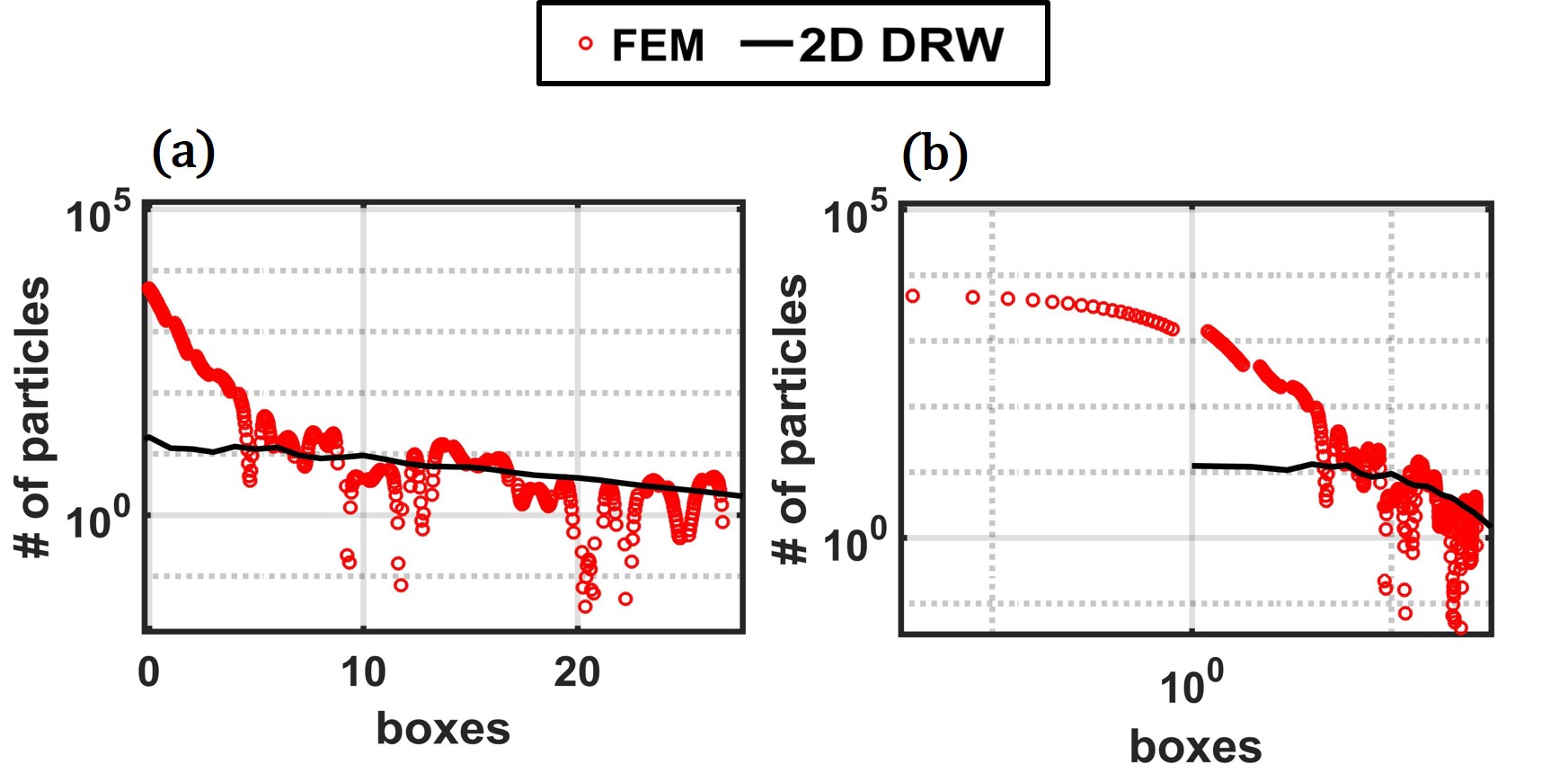}}
  \caption{Comparison between the FE and DRW numerical results of the intensity distribution for a 2D periodic square lattice of scatterers with a monopole acoustic source at its center. (a) semilogarithmic and (b) logarithmic scale plot.}
\label{Fig_20}
\end{figure}

The comparison between the intensity distributions obtained by numerical FE simulations and by equivalent 2D DRW model is reported in Fig.~\ref{Fig_20}.  

Also in this case, the DRW model is in very good agreement with the FE simulations and, most important, is perfectly capable of capturing the anomalous (power-law) decay of the tails of the distribution. This result provides the conclusive proof that the anomalous behavior observed in the square lattice is in fact the result of long-range (L{\'e}vy flights) interactions due to scattering events occurring along different directions of propagation that are characterized by anisotropic dispersion.

\section{$\alpha$-stable distributions and fractional diffusion equation}  \label{fractional diffusion}

The renormalization criterion used in section \S \ref{Numerical_1} to determine the existence of the anomalous diffusion regime is theoretically well-grounded but it does not allow a convenient approach to classify the anomalous regime. This classification typically requires the analysis of the time scales involved in the evolution of the moments of the distribution \cite{metzler2000random}. Here we suggest a different approach that, not only provides a more direct classification based on the available data, but opens new routes for an analytical treatment of the resulting diffusion problem.

The intensity distributions reported in Fig.~\ref{Fig_14} suggest a power-law behavior of the tails. Recent studies \cite{mainardi1996fractional,mainardi1995fractional, benson2000application,benson2001fractional} have shown that, for physical phenomena exhibiting this characteristic distribution of the field variables, the governing equations are generalizations to the fractional order of the classical diffusion equation. Power-law distributions, associated with infinite variance random variables (the so called L{\'e}vy flights), are in the domain of attraction of $\alpha$-stable random variables also called L{\'e}vy stable densities (their properties are summarized in Appendix \ref{Appendix}). On the other hand finite-variance random variables are in the Normal domain of attraction that is a subset of L{\'e}vy stable densities. This suggests that the trend of the tails carries information about the $\alpha$-stable order of the underlying distribution.

In order to show that this situation occurs also in the present case, we performed numerical fits of the acoustic intensity profiles (Fig.~\ref{Fig_14}) using $\alpha$-stable distributions. 

The four parameters defining the $\alpha$-stable distributions were obtained by numerically solving a nonlinear optimization problem. The most important parameter is the characteristic exponent $\alpha$ (also called the index of stability) that is also connected to the slope of the tails. For the square lattice distribution, the values of $\alpha$ determined with the optimization procedure are $\alpha = 0.89$ and $\alpha = 0.57$ for the passband and bandgap excitation wavelengths, respectively.

\begin{figure}[h!]
\center{\includegraphics[width=\linewidth]{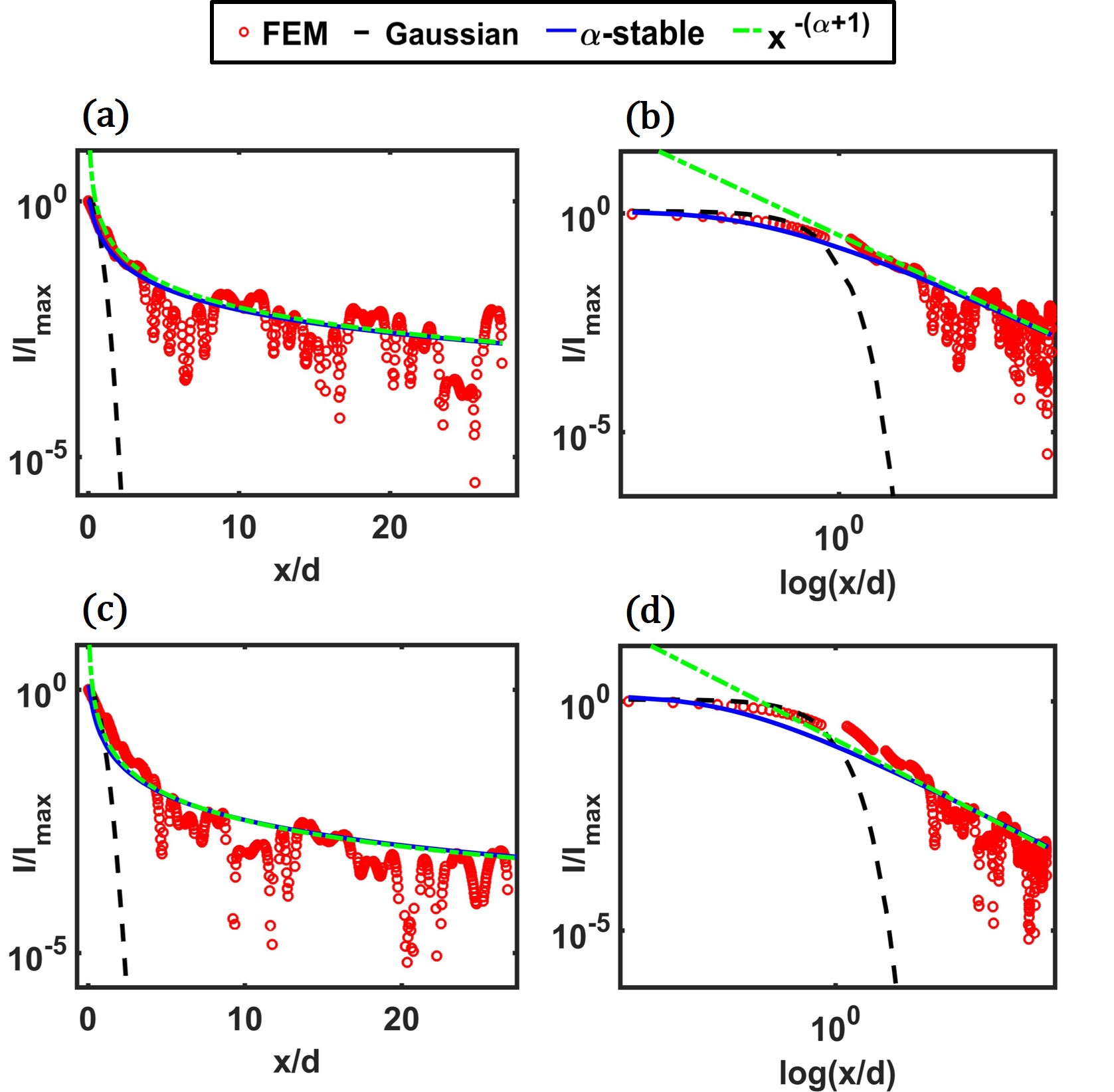}}
  \caption{Gaussian and $\alpha$-stable fits of the acoustic intensity versus the non-dimensional distance from the source. (a) semilogarithmic and (b) logarithmic plot for $\lambda/d = 2.3352$. (c) semilogarithmic and (d) logarithmic plot for $\lambda/d = 2.1552$.}
\label{Fig_21}
\end{figure}

In order to show that the order of the $\alpha$-stable distribution effectively describes the anomalous diffusive dynamics of the system, we use a generalized fractional diffusion equation \cite{mainardi2007fundamental}:

\begin{align} \label{Fractional diffusion}
\begin{split}
_{x}\textrm{D}_{\theta}^{\alpha}u(x,t)=_{t}\textrm{D}_{*}^{\beta}u(x,t) \quad x \in \mathbb{R} , \quad  t \in \mathbb{R}^+
\end{split}
\end{align}

\noindent where $\alpha$, $\theta$, $\beta$ are real parameters always restricted as follows:

\begin{align} \label{Restrictions}
\begin{split}
0< \alpha \leq2, \quad |\theta|\leq min\left \{ \alpha,2-\alpha \right \}, 0 <\beta \leq 2.
\end{split}
\end{align}

In Eq.~(\ref{Fractional diffusion}), $u = u(x,t)$ is the field variable, $_{x}\textrm{D}_{\theta}^{\alpha}$ is the \textit{Riesz-Feller} space fractional derivative of order $\alpha$, and $_{t}\textrm{D}_{*}^{\beta}$ is the Caputo time-fractional derivative of order $\beta$.  The fractional operator in this equation exhibits a non-local behavior which makes it ideally suited to model dynamical systems dominated by long-range interactions.

Mainardi \cite{mainardi2007fundamental} reported the Green's function for a Cauchy problem based on the space-time fractional diffusion equation. The self-similar nature of the solution allows the application of a similarity method that separates the solution into a space dependent (the reduced Green's function $K$) and a time dependent term. In our system, we use a harmonic (constant amplitude) source and we analyze the steady state response, that is we consider a self-similar problem. In other terms, the reduced Green's function $K$ proposed by Mainardi coincides with the normalized solution of the forced fractional diffusion equation governing our problem: 

\begin{align} \label{Reduced Green function}
\begin{split}
\textrm{K}_{\alpha,\beta}^{\alpha}(x) = \frac{1}{\pi x}\sum_{n=1}^{\infty}\frac{\Gamma (1+\alpha n)}{\Gamma(1+\beta n)}\sin\left [ \frac{n\pi}{2}(\theta-\alpha) \right ](-x^{-\alpha})^n.
\end{split}
\end{align}
Note that this solution is valid in the case $\alpha<\beta$. In our case, $\beta=1$ to model a space fractional diffusion equation. 

\begin{figure}[h!]
\center{\includegraphics[width=\linewidth]{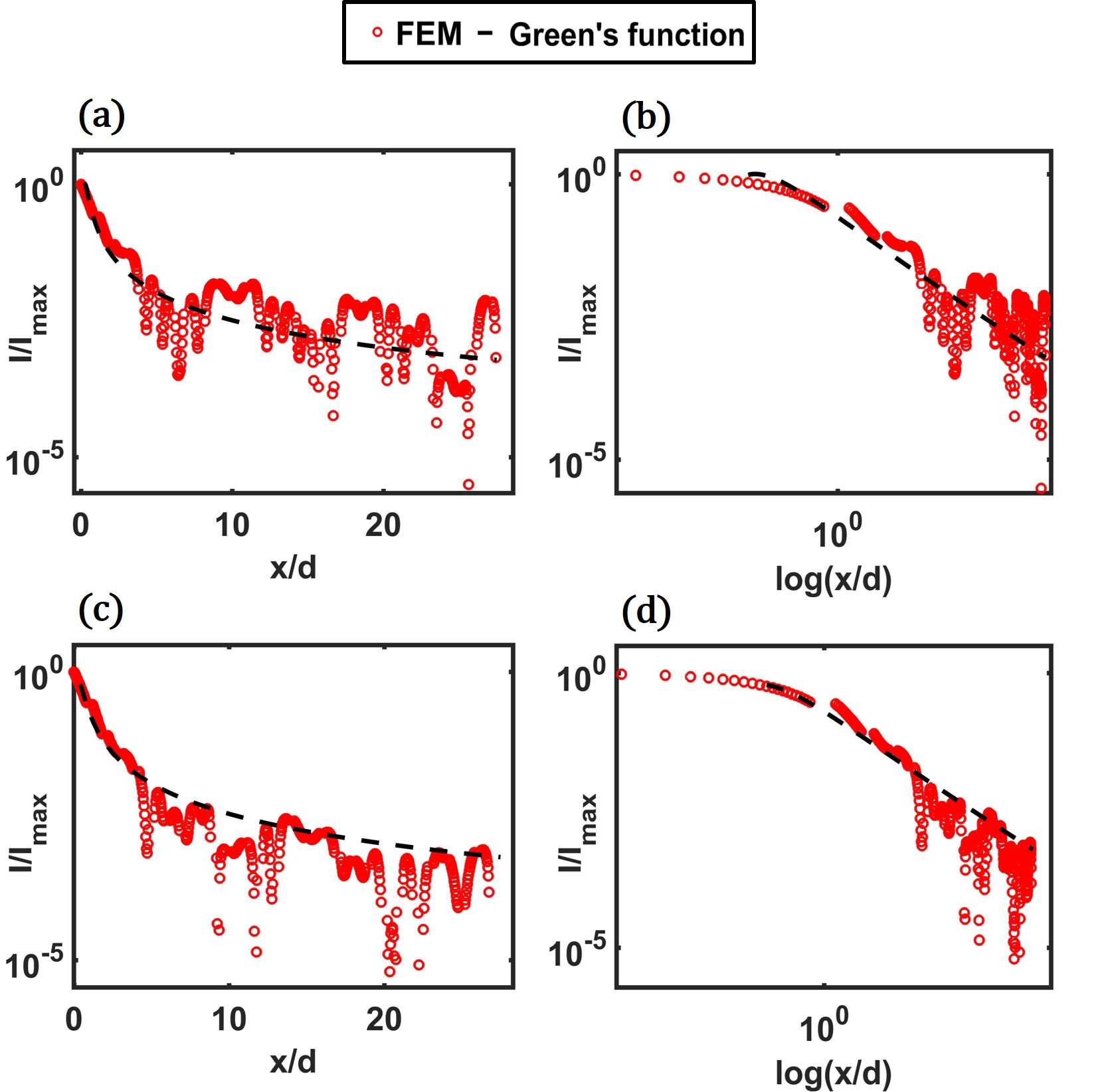}}
  \caption{Comparison of the intensity distribution with the reduced Green's function solution of the space-fractional diffusion equation for the non-dimensional wavelength $\lambda/d = 2.3352$, in (a) semilogarithmic and (b) logarithmic scales and the non-dimensional wavelength $\lambda/d = 2.1552$, in (c) semilogarithmic and (d) logarithmic scales.}
\label{Fig_22}
\end{figure}

Fig.~\ref{Fig_22} shows the comparison between the normalized acoustic intensity from the FE numerical data and the result from the reduced Green's function $K$ (Eq.~(\ref{Reduced Green function})) calculated for the order $\alpha$ obtained by the previous $\alpha$-stable fits.
 
The above results clearly show that the fractional diffusion equation is able to capture the heavy-tailed behavior of the intensity distribution with good accuracy. They also confirm that the use of $\alpha$-stable fits provides a reliable approach to classify the anomalous behavior and to extract the corresponding fractional order of the operator.
The above numerical results provide also further confirmation that the observed dynamic behavior from the full field numerical simulations is in fact dominated by anomalous diffusion. 
These results are particularly relevant if seen in a perspective
of developing predictive capabilities for transport processes in highly inhomogeneous systems. As an example, fractional models would provide an excellent framework for the solution of inverse problems in imaging and remote sensing through highly scattering media. The
ability to properly capture a mixed transport behavior, such as partially propagating and diffusive, would allow extracting more information from the measured response therefore improving the sensitivity and resolution of these approaches.
From a broader perspective, this methodology has general applicability and could be extended to a variety of applications involving wave-like
field transport such as those mentioned in the introduction.

\section{Conclusions}   \label{Conclusion}

In this paper, we investigated the scattering behavior of sound waves in a perfectly periodic acoustic medium composed of a square lattice of hard cylinders in air. From a general perspective, the most remarkable result consists in the observation of the occurrence of anomalous hybrid transport in perfectly periodic lattice structures without disorder or random properties. This result is particularly relevant because the anomalous response of a scattering system was previously observed only in systems with either stochastic material or geometric properties.
By using a combination of theoretical and numerical models, both deterministic and stochastic, it was determined that the existence of long-range interactions associated with the anisotropy of the dispersion bands was the driving factor leading to the occurrence of the anomalous transport behavior. The resulting diffused intensity fields were characterized by heavy-tails with marked asymptotic power-law decay, that were well described by $\alpha$-stable distributions. It was also shown that the $\alpha$-stable nature of the dynamic response provided a reliable approach for the classification and characterization of the non-local effect via the intrinsic parameters of $\alpha$-stable distributions. 

Observing that $\alpha$-stable distributions represent the fundamental kernel for the solutions of fractional continuum models, we showed that a space fractional diffusion equation having the order predicted by the $\alpha$-stable fit of the acoustic intensity was capable of capturing very accurately the characteristic features of the anomalous transport process. From a general perspective, this approach can be interpreted as a fractional order homogenization of the periodic medium which is capable of mapping the complex inhomogeneous system to a (fractional) governing equation that still accepts an analytical solution.

This latter characteristic is particularly remarkable if seen from a practical application perspective because it could open the way to accurate and non-iterative inverse problems that play a critical role in remote sensing, imaging, and material design, just to name a few.
Another key observation concerns the strong deviation of the tails of the acoustic intensity from the Gaussian distribution which highlights that much information is still contained in the tails. This aspect is particularly relevant for imaging and sensing in scattering media because traditional analytical methodologies typically assume a Gaussian distribution of the measured intensity field hence leading to two main drawbacks: 1) the loss of important information about the internal structure of the medium which is contained in the tails, and 2) the lack of a proper model capable of extracting and interpreting this information from measured data.

\section*{Acknowledgments}
The authors gratefully acknowledge the financial support of the National Science Foundation under the grants DCSD CAREER $\#1621909$ and of the Air Force Office of Scientific Research under Grant No. YIP FA9550-15-1-0133.

\appendix*
\section{Appendix A}   \label{Appendix}

This appendix summarizes some basic properties of the $\alpha$-stable distributions that have been used to analyze and interpret the simulation data in the paper.
The family of $\alpha$-stable distributions are defined by the Fourier transform of their characteristics functions $\psi(w)$ that can be  written in the explicit form as\cite{herranz2004alpha,benson2002fractional}:
\begin{eqnarray} \label{alpha-stab0}
\psi(w)=\exp\left \{ i\mu w-\gamma\left | w \right |^\alpha B_{w,\alpha} \right \} \\
B_{w,\alpha}=\left\{\begin{matrix}
\left [1+i\beta \operatorname{sgn}(w) \tan\frac{\alpha \pi}{2} \right ]  \quad  \alpha\neq1 \nonumber \\ 
\left [1+i\beta \operatorname{sgn}(w) \frac{2}{\pi} \log\left | w \right | \right ]  \quad  \alpha=1 
\end{matrix}\right.
\end{eqnarray}

where $0<\alpha\leq 2$, $-1\leq\beta\leq 1$, $\gamma>0$, and $-\infty<\mu<\infty$. The parameters $\alpha$, $\beta$, $\gamma$ and $\mu$ uniquely and completely identify the stable distribution.

\begin{enumerate}
  \item The parameter $\alpha$ is the \textit{characteristic exponent}, or the \textit{stability parameter}, and it defines the degree of impulsiveness of the distribution. As $\alpha$ decreases the level of impulsiveness of the distribution increases. For $\alpha=2$ we recover the Gaussian distribution. A particular case is obtained for $\alpha=1$ and $\beta=0$ that corresponds to the Cauchy distribution. For $\alpha \notin (0,2]$ the inverse Fourier transform $\psi(w)$ is not positive-definite and hence is not a proper probability density function.
  
   \item The parameter $\beta$ is the \textit{symmetry}, or \textit{skewness parameter}, and determines the skewness of the distribution. Symmetric distributions have $\beta=0$, whereas $\beta=1$ and $\beta=-1$ correspond to completely skewed distributions.
   
   \item The parameter $\gamma$ is the \textit{scale parameter}. It is a measure of the spread of the samples from a distribution around the mean.
   
   \item The parameter $\mu$ is the  \textit{location parameter} and corresponds to a shift in the $x$-axis of the pdf. For a symmetric $(\beta=0)$ distribution, $\mu$ is the mean when $1<\alpha\leq 2$ and the median when $0<\alpha\leq 1$.
   
\end{enumerate}

The characteristic functions described in Eq.~(\ref{alpha-stab0}) are equivalent to a probability density function and do not have analytical solutions except for few special cases. The main feature of these characteristic functions is the presence of heavy-tails when compared to a Gaussian distribution. The probability density functions with tails heavier than Gaussian are also denoted as \textit{impulsive}. An impulsive process is characterized by the presence of large values that significantly deviates from the mean value of the distribution with non-negligible probability. In this sense the $\alpha$-stable distribution represents a generalization of the Gaussian distribution that allows to model impulsive processes by using only four parameters instead of an infinite number of moments. The possibility of describing the distribution of particles in anomalous diffusion phenomena by using $\alpha$-stable distributions has numerous advantages: 1) many methods exist to perform statistical inference on $\alpha$-stable environments \cite{nikias1995signal, janicki1993simulation}, 2) these distributions are simple because they are completely characterized by only four parameters, 3) the use of $\alpha$-stable distributions finds a theoretical justification in the fact that they satisfy the generalized central limit theorem which states that the limit distribution on infinitely many {i.i.d.} random variables, is a stable distribution, 4) they include the Gaussian distribution as a particular case for a specific set of parameters. These distributions are stable since the output of a linear system in response to $\alpha$-stable inputs is again $\alpha$-stable.

\bibliographystyle{unsrt}
\bibliography{ref}

\end{document}